\documentclass[prd,superscriptaddress,showpacs,preprint]{revtex4}
\usepackage{chay,epsfig}
\begin{document}
\title{Soft Wilson lines in  soft-collinear effective theory}
\author{Junegone Chay}\email{chay@korea.ac.kr} 
\affiliation{Department of Physics, Korea University, Seoul 136-701,
Korea}

\author{Chul Kim}
\affiliation{Department of
  Physics and Astronomy, University of Pittsburgh, PA 15260, U.S.A.}

\author{Yeong Gyun Kim}
\affiliation{Department of Physics, Korea University, Seoul 136-701,
Korea}

\author{Jong-Phil Lee}
\affiliation{Department of Physics, Yonsei University, Seoul
  120-749, Korea}
\preprint{KUPT-04-02}
\begin{abstract}
The effects of the soft gluon emission in hard scattering processes
at the phase boundary are resummed in the soft-collinear effective
theory (SCET). In SCET, the soft gluon emission is decoupled from the
energetic collinear part, and is obtained by the vacuum expectation
value of the soft Wilson-line operator. The form of the soft Wilson
lines is universal in deep inelastic scattering, in the Drell-Yan
process, in the jet production from $e^+e^-$ collisions, and in the
$\gamma^* \gamma^* \rightarrow \pi^0$ process, but its analytic
structure is slightly different in each process. The anomalous
dimensions of the soft Wilson-line operators for these processes are
computed along the light-like path at leading order in SCET and to
first order in $\alpha_s$, and the renormalization group behavior of
the soft Wilson lines is discussed.   
\end{abstract}

\maketitle

\section{Introduction}
The factorization theorem \cite{css} has been one of the most
important issues in QCD. It states that scattering cross sections or
amplitudes can be written as a product or a convolution of the hard,
the collinear, and the soft parts, and each part depends on a single
scale. The factorization of the hard, collinear, and soft
parts in various processes was studied previously in the 
full theory, considering the appropriate kinematic regions in which
the particles involved can be either collinear or soft
\cite{collins89,grammer73,Korchemskaya:1992je,
  Korchemsky:1992xv,Korchemsky:1993uz,Korchemsky:1994is}. 
It has been a complicated task to disentangle the contributions from
all the possible kinematic regions, but the soft-collinear effective
theory (SCET)
\cite{Bauer:2000ew,Bauer:2000yr,Bauer:2001ct,Bauer:2001yt} can be
helpful in understanding this difficult procedure since SCET is
formulated in such a way that the collinear interactions and the soft
interactions are decoupled to all orders in $\alpha_s$ from the
beginning.

SCET successfully describes the factorization properties in $B$ decays
such as $B\rightarrow D\pi$ \cite{Bauer:2001cu}, $B\rightarrow \gamma
e\nu$ \cite{Lunghi:2002ju}, nonleptonic
$B$ decays \cite{Chay:2003zp,Chay:2003ju,Bauer:2004tj} and
$B\rightarrow K^* \gamma$ \cite{Chay:2003kb}. And it can be applied to
other high-energy processes to probe the factorization properties
\cite{Fleming:2004rk}. In order to show the utility of SCET in
high-energy processes, we will consider deep inelastic scattering, the
Drell-Yan process, jet production from $e^+e^-$ collisions, and the
$\pi$-$\gamma$ form factor with two virtual photons, where all the
hadrons in these processes are light and energetic so that we can
apply SCET. The treatment of these high-energy processes with SCET has
been extensively discussed in Ref.~\cite{Bauer:2002nz}.

In SCET the factorization occurs in each 
process, in which the scattering cross section or the form factor can
be written as a convolution of the hard part (the Wilson coefficients
from matching the full theory onto $\mathrm{SCET}_{\mathrm{I}}$), the
collinear part (the jet functions from matching
$\mathrm{SCET}_{\mathrm{I}}$ onto $\mathrm{SCET}_{\mathrm{II}}$), and
the soft part (the matrix elements of the remaining operators in
$\mathrm{SCET}_{\mathrm{II}}$). The matrix elements between hadronic
states are  nonperturbative and cannot be 
computed from first principles, but their scaling behavior can be
obtained using perturbation theory. In high-energy scattering processes,
the matrix elements of gauge-invariant operators are usually
parameterized as either the parton distribution functions for hadrons
in the initial states, or the fragmentation functions for the
final-state hadrons \cite{collins82}.    

In this paper, we are mainly interested in the nonperturbative part
from the soft interactions in SCET. The effect of soft interactions in
the soft part does not usually appear at leading order in SCET since
it cancels in many processes. However, near the boundary of 
the phase space, the cancellation is incomplete and the
nonperturbative effect of soft gluon emission can be important. In
this case, the conventional operator product expansion breaks down,
but still we can consider gauge-invariant nonlocal operators which are
connected by an operator with a string of gauge fields. The endpoint
behavior in deep inelastic scattering using SCET was first discussed
in Ref.~\cite{Manohar:2003vb}, focusing on the scaling behavior of the
matrix elements of the collinear operators. The nonperturbative
effects of the soft gluon emission in $e^+e^-$ collisions were first
considered in terms of SCET in
Refs.~\cite{Bauer:2002ie,Bauer:2003di}. 

Near the phase boundary, the final operators, of which the matrix
elements between hadronic states describe nonperturbative effects,
consist of collinear fields, which are connected by soft Wilson
lines. In SCET, since the soft interactions are decoupled from the
collinear sector, we can decompose the operators in terms of the
collinear operators and the soft Wilson lines and we can consider the
matrix elements of these two kinds of operators. Though the matrix 
element of the soft Wilson lines cannot be computed in perturbation
theory, the renormalization group behavior of the soft Wilson-line
operator can be derived.

There has been such consideration in the full theory. For example, in
Ref.~\cite{Korchemsky:1993uz}, the authors considered the vacuum 
expectation value of a soft Wilson line which is defined as
\begin{equation} \label{wc}
W (C) \equiv \frac{1}{N} \langle 0| P \exp \Bigl( ig \oint_C dz_{\mu}
A^{\mu} (z) \Bigr) |0\rangle.
\end{equation}
In Ref.~\cite{Korchemskaya:1992je}, the vacuum
average of a soft Wilson loop was analyzed, which is given by
\begin{equation} \label{wtc}
  W_T(C) =\frac{1}{N} \langle 0|\mathrm{tr}  TP \exp \Bigl( ig \oint_C
  dz_{\mu} A^{\mu} (z)   \Bigr) |0\rangle.
\end{equation}
Here the integration path $C$ is determined by the kinematics of the
processes, $T$ orders gauge fields $A_{\mu}^a (z)$ in time, and $P$
orders the generators $T^a$ of the $SU(N)$ gauge group along the path
$C$. The difference between $W(C)$ and  $W_T (C)$ is that the
gluon fields in $W(C)$ are ordered along the path $C$, but not according to
time. Therefore on different parts of the path $C$, the gluon fields
are time or anti-time ordered. The minute difference in the
definitions of $W_T (C)$ and $W(C)$ affects the analytic structure of
the soft Wilson lines.

The main theme of the paper is to study the analytic structure of the 
Wilson lines, especially the soft Wilson lines in SCET near the
boundary of the phase space. We show that the soft Wilson lines appear
universally in deep inelastic scattering, in the Drell-Yan process and
in the jet production from $e^+e^-$ collisions, in which the Wilson
lines appear in the matrix elements squared or the discontinuity of
the forward scattering amplitude. They also appear in the
$\pi$-$\gamma$ form factor, in which the soft Wilson line appears in
an amplitude. After we identify the analytic structure of the soft
Wilson lines, we compute the anomalous dimensions of the soft Wilson
lines in SCET to see the renormalization group behavior.

The conventional approach \cite{css} for this analysis is to separate
the region of momentum and to extract the contribution from the
collinear region and the soft region, but the advantage of SCET comes
from the fact that this separation is performed automatically when we
perform the two-step matching \cite{Bauer:2002aj} with the two
effective theories called $\mathrm{SCET}_{\mathrm{I}}$ and
$\mathrm{SCET}_{\mathrm{II}}$. Near the boundary of the phase space,
the intermediate states can have momentum of order $p_X^2 \sim Q^2
(1-x)$, where $Q$ is the large momentum scale with  $x\sim 1$, but
still $p_X^2 \gg \Lambda_{\mathrm{QCD}}^2$. First, we integrate the
degrees of freedom of order $p^2 \sim Q^2$ from the full theory and
match onto the intermediate effective theory,
$\mathrm{SCET}_{\mathrm{I}}$. Here the ultrasoft (usoft) particles
with momentum of order $\Lambda \sim \Lambda_{\mathrm{QCD}}$ can
interact with collinear particles. Then we integrate out the degrees
of freedom of order $p^2 \sim Q^2 (1-x)$ to go down to
$\mathrm{SCET}_{\mathrm{II}}$. In $\mathrm{SCET}_{\mathrm{II}}$, the
soft particles and the collinear particles are decoupled, and the
effects of soft gluon emission in SCET can be studied without regard
to the collinear sector. The soft Wilson line operator in SCET, which
we will elaborate in detail, takes the form
\begin{equation} \label{keta}
  K(\eta) = \frac{1}{N}\mathrm{tr} \Bigl( \overline{S}^{\dagger} S
  \delta (\eta +in\cdot \partial) S^{\dagger} \overline{S} \Bigr),
\end{equation}
where the Wilson lines $S$ and $\overline{S}$ are the Fourier
transforms of
\begin{equation}
  S(z) = \exp \Bigl[ ig \int ds n\cdot A_s
  (ns+z) \Bigr], \ \overline{S} (z)=  \exp \Bigl[ ig \int ds
  \overline{n}\cdot A_s   (\overline{n}s+z) \Bigr],
\end{equation}
and here we suggest how to prescribe the path ordering, which is 
determined by the processes under consideration. We also study the
renormalization properties of the soft Wilson-line operators for a
path $C$ lying on the light cone in SCET after we separate the soft
interactions in $\mathrm{SCET}_{\mathrm{II}}$. This is in contrast to
Eqs.~(\ref{wc}) and (\ref{wtc}) since the paths are partially lying on
the light-cone. In our approach, we put all the energetic
collinear particles on the light-cone and consider the soft Wilson
lines on these light cones at leading order in SCET.

The structure of the paper is as follows: In Sec.~\ref{sec2}, we study
in detail the analytic structure of the soft Wilson lines in SCET.
In Sec.~\ref{sec3}, we explain how the operator
$K(\eta)$ for the soft gluon emission appears in deep inelastic
scattering, in the Drell-Yan process, in the jet production from $e^+
e^-$ scattering, and in the $\pi$-$\gamma$ form factor with two
virtual photons. In
Sec.~\ref{sec4}, we compute the radiative
correction for $K(\eta)$ at one loop and derive the renormalization
group equation. We also show that the result is consistent with the
result obtained by Korchemsky et
al. \cite{Korchemskaya:1992je,Korchemsky:1992xv,Korchemsky:1993uz}. In
the final section, we present conclusions and compare the conventional
approach with the approach in SCET.

\section{Analytic structure of the Wilson lines\label{sec2}}
\begin{figure}[b]
\begin{center}
\epsfig{file=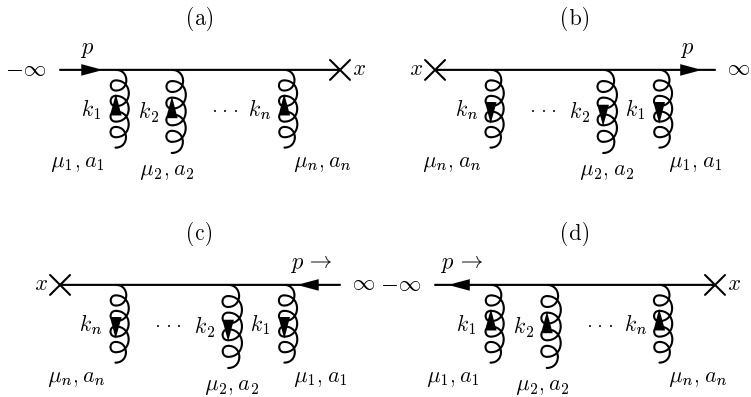, width=10.0cm}
\end{center}
\vspace{-0.5cm}
\caption{Attachment of soft gluons to (a) an incoming quark, (b) an
  outgoing quark, (c) an outgoing antiquark, and (d) an incoming
  antiquark.}
\label{figwilson}
\end{figure}

The analytic structure of the soft Wilson lines is important in
computing radiative corrections because the position of the poles is
determined by the appropriate $i\epsilon$ prescription. In SCET, the
soft Wilson line is obtained by factorizing
the usoft interactions in $\mathrm{SCET}_{\mathrm{I}}$. One way of
deriving the Wilson line is to attach usoft gluons to collinear fields
and use the eikonal approximation. There are four possible cases and
they are shown in Fig.~\ref{figwilson}.

In Fig~\ref{figwilson} (a), soft gluons are attached to
an incoming quark from $-\infty$ to $x$ or an outgoing antiquark from
$x$ to $-\infty$,  and the Wilson line is given by
\begin{equation}
  Y=1+\sum_{m=1}^{\infty} \sum_{\mathrm{perms}} \frac{(-g)^m}{m!}
  \frac{n\cdot A_s^{a_n} \cdots n\cdot A_s^{a_1}}{n\cdot
  \Bigl(\sum_{i=1}^n k_i +i\epsilon\Bigr) \cdots (n\cdot k_1
  +i\epsilon)}   T_{a_n} \cdots T_{a_1},
\end{equation}
which is written as
\begin{equation} \label{yex}
Y = \sum_{\mathrm{perm}} \exp \Bigl[  \frac{1}{n\cdot
  \mathcal{P}+i\epsilon} (-gn\cdot A_s)   \Bigr],
\end{equation}
where $\mathcal{P}^{\mu}$ is the momentum operator. Here $A_s^{\mu}$
denotes the usoft gluon in $\mathrm{SCET}_{\mathrm{I}}$. In the
literature, this is commonly denoted as $A_{us}^{\mu}$, and after we
go down to $\mathrm{SCET}_{\mathrm{II}}$, we relabel them as the soft
gluon $A_s^{\mu}$. However, we will use $A_s^{\mu}$ for the usoft
(soft) gluon in $\mathrm{SCET}_{\mathrm{I}}$
($\mathrm{SCET}_{\mathrm{II}}$) for simplicity. The exponentiated form
$Y$ is related to the Fourier transform $Y(x)$ of the path-ordered
exponential 
\begin{equation}
\label{yfo}
  Y(x) = P \exp \Bigl(ig \int_{-\infty}^x ds n\cdot A_s (ns) \Bigr),
\end{equation}
where the path ordering $P$ means that the fields are ordered in such
a way that the gauge fields closer (farther) to the point $x$ are
moved to the left (right).

Note that the Feynman ``$i\epsilon$'' prescription enforces the path
ordering. This can be seen if we consider the Fourier transform of the
exponent in Eq.~(\ref{yex}). $n\cdot A_s (x)$ depends on the coordinate
$x$, but since the usoft gluons are attached to the collinear particle
moving in the $n^{\mu}$ direction, $n\cdot A_s (x)$ depends only on
$\overline{x} =\overline{n}\cdot x/2$ at leading order in
$\Lambda$. It means that the Fourier transform of $n\cdot A_s (x)$
depends only on $n\cdot q$. Therefore we can consider the
one-dimensional Fourier transform
\begin{equation}
  n\cdot A_s (x) = \frac{1}{2\pi} \int dn\cdot q e^{-in\cdot q
  \bar{x}} n\cdot A_s (n\cdot q), \  n\cdot A_s (n\cdot q) = \int
  d\overline{x} e^{in\cdot q \bar{x}} n\cdot A_s (x),
\end{equation}
since the remaining components of the Fourier transform yield only the
 delta functions.
The Fourier transform of the  exponent in Eq.~(\ref{yex}) can be
 written as 
\begin{equation} \label{yf}
  \frac{-g}{2\pi} \int dn\cdot q \frac{e^{-in\cdot q
  \overline{x}}}{n\cdot q +i\epsilon} n\cdot A_s (n\cdot q) 
= \frac{-g}{2\pi} \int_{-\infty}^{\infty}
 d\overline{y} \int dn\cdot q \frac{e^{in\cdot
  q(\overline{y} -\overline{x})}}{n\cdot q+i\epsilon} n\cdot A_s
  (\overline{y}).
\end{equation}
We can perform the integration over $n\cdot q$ in the complex 
plane. Because the pole is in the lower half plane, the integration
over $n\cdot q$ becomes $-2\pi i  \theta (\overline{x}
-\overline{y})$. And Eq.~(\ref{yf}) is given by
\begin{equation}
 ig  \int_{-\infty}^{\overline{x}}
   d\overline{y} n\cdot A_s(\overline{y}).
\end{equation}

In order to see how the path ordering is specified, let us consider
the Fourier transform of Eq.~(\ref{yex}) at order $g^2$, which
is given by 
\begin{eqnarray} \label{2nds}
&&\frac{1}{n\cdot \mathcal{P} +i\epsilon} gn\cdot A_s 
  \frac{1}{n\cdot \mathcal{P} +i\epsilon} g n\cdot A_s 
 \\
&&\rightarrow \int
  \frac{dn\cdot q_1 dn\cdot q_2}{(2\pi)^2} e^{-i n\cdot (q_1 +q_2)
  \bar{x}} \frac{gn\cdot A_s (n\cdot
  q_2) gn\cdot A_s (n\cdot q_1)}{\Bigl(n\cdot (q_1 +q_2) +i\epsilon
  \Bigr)(n\cdot   q_1 +i\epsilon)} \nonumber \\
&&=\int d\overline{y} d\overline{z}  \int
  \frac{dn\cdot q_1 dn\cdot q_2}{(2\pi)^2} e^{i n\cdot q_1 (\bar{y}
  -\bar{x})} e^{in\cdot q_2 (\bar{z} -\bar{x})}
 \frac{gn\cdot A_s (\overline{z}) gn\cdot A_s (\overline{y})}{\Bigl(
  n\cdot   (q_1 +q_2) +i\epsilon \Bigr)(n\cdot   q_1 +i\epsilon)}
  \nonumber \\ 
&&= \int d\overline{y} d\overline{z} gn\cdot A_s (\overline{z})
  gn\cdot A_s (\overline{y})  \int \frac{dn\cdot q_1}{2\pi}
  \frac{ e^{i n\cdot q_1 (\bar{y}
  -\bar{x})}}{n\cdot q_1 +i\epsilon} \int \frac{dn\cdot q_2}{2\pi}
  \frac{e^{in\cdot q_2 (\bar{z} -\bar{x})}}{n\cdot (q_1 +q_2)
  +i\epsilon} \nonumber \\
&&=  \int  d\overline{y} d\overline{z} gn\cdot A_s (\overline{z})
  gn\cdot A_s (\overline{y})  \int \frac{dn\cdot q_1}{2\pi}
  \frac{ e^{i n\cdot q_1 (\bar{y}
  -\bar{x})}}{n\cdot q_1 +i\epsilon} (-i) e^{-i n\cdot q_1 (\bar{z}
  -\bar{x})} \theta (\overline{x} -\overline{z}) \nonumber \\
&&= (-i)^2   \int d\overline{y} d\overline{z} gn\cdot A_s (\overline{z})
  gn\cdot A_s (\overline{y})  \theta (\overline{x} -\overline{z})
  \theta (\overline{z} -\overline{y}) \nonumber \\
&&= (-i)^2 \int_{-\infty}^{\bar{x}} d\overline{z} gn\cdot A_s
  (\overline{z}) \int_{-\infty}^{\bar{z}} d\overline{y}  gn\cdot A_s
  (\overline{y}) =\frac{(-i)^2}{2!} \int_{-\infty}^{\bar{x}}
  d\overline{z} \int_{-\infty}^{\bar{x}} d\overline{y} P[gn\cdot A_s
  (\overline{z}) gn\cdot A_s (\overline{y})]. \nonumber
\end{eqnarray}
In Eq.~(\ref{2nds}), the integrations over $n\cdot q_1$ and $n\cdot
q_2$ are performed in the complex plane, and the final relation with
the path ordering is obtained by changing the region of
integration. Similarly, for the $m$-th order term, we have the
relation
\begin{equation}
  \Bigl[ \frac{1}{n\cdot \mathcal{P} +i\epsilon} gn\cdot A_s
  \Bigr]^m \rightarrow \frac{(-i)^m}{m!} \int_{-\infty}^{\bar{x}}
  d\overline{y}_1 \cdots d\overline{y}_m P\Bigl[ gn\cdot A_s
  (\overline{y}_1) \cdots gn\cdot A_s (\overline{y}_m) \Bigr],
\end{equation}
which shows that Eq.~(\ref{yex}) indeed specifies the path-ordered
exponential in Eq.~(\ref{yfo}).

We can find an equivalent expression for $Y$ if we replace the incoming
gluons by the outgoing gluons in Fig.~\ref{figwilson} (a). Then the
relative sign of the quark momentum $p$ and the gluon momenta $k_i$
becomes opposite and $Y$ can be written as
\begin{equation}
  Y= \sum_{\mathrm{perm}}\exp \Bigl[-g n\cdot A_s \frac{1}{-n\cdot
  \mathcal{P}^{\dagger}+i\epsilon}   \Bigr],
\end{equation}
but the corresponding Fourier transform is given by the same form in
Eq.~(\ref{yfo}).

We can proceed in a similar way for other diagrams. Fig.~\ref{figwilson}
(b) describes an outgoing quark from $x$ to $\infty$, or an incoming
antiquark from $\infty$ to $x$ emitting soft gluons, and the
soft Wilson line, and its Fourier transform are written as
\begin{equation}
  \tilde{Y}^{\dagger} = \sum_{\mathrm{perm}} \exp \Bigl[ -gn\cdot A_s
  \frac{1}{n\cdot   \mathcal{P}^{\dagger} +i\epsilon}\Bigr], \
\tilde{Y}^{\dagger} (x) = P \exp \Bigl(ig\int_{x}^{\infty} ds n\cdot
  A_s   (ns) \Bigr).
\end{equation}
Our notation is such that the Wilson lines with a tilde propagate
between $x$ and $\infty$, and the Wilson lines without a tilde
propagate between $-\infty$ and $x$. Fig.~\ref{figwilson} (c) is the
case in which an outgoing antiquark is
moving from $x$ to $\infty$, or an incoming quark from $\infty$ to
$x$, emitting soft gluons, and the corresponding Wilson line
is given by
\begin{equation}
  \tilde{Y} = \sum_{\mathrm{perm}}\exp \Bigl[\frac{1}{n\cdot
    \mathcal{P}-i\epsilon} (-gn\cdot A_s) \Bigr], \
\tilde{Y} (x) = \overline{P} \exp \Bigl(-ig
    \int_x^{\infty} ds n\cdot   A_s (ns)\Bigr),
\end{equation}
where $\overline{P}$ means anti-path ordering. 
Fig.~\ref{figwilson} (d) describes an incoming antiquark from
$-\infty$ to $x$, or an outgoing quark from $x$ to $-\infty$ emitting
soft gluons, and the Wilson line is given as
\begin{equation}
 Y^{\dagger} = \sum_{\mathrm{perm}}\exp \Bigl[ -gn\cdot A_s
  \frac{1}{n\cdot
  \mathcal{P}^{\dagger}   -i\epsilon} \Bigr], \
Y^{\dagger} (x) = \overline{P} \exp \Bigl(
  -ig\int_{-\infty}^x ds    n\cdot A_s (ns) \Bigr).
\end{equation}
From now on, we will omit the summation over all the possible
permutations in $Y$, $Y^{\dagger}$, $\tilde{Y}$ and
$\tilde{Y}^{\dagger}$. These results are summarized in
Table~\ref{table1}. And we 
explicitly construct the Wilson lines to satisfy the unitarity
$Y^{\dagger} (x) Y(x)= \tilde{Y}^{\dagger} (x) \tilde{Y} (x)=1$
and it is also true for $\overline{Y}$ and
$\tilde{\overline{Y}}$. With our notation, $Y \tilde{Y}^{\dagger}$
does not have to be 1 unless we fix the gauge as
$n\cdot A_s=0$, but in this case, $Y$, $\tilde{Y}$ become formally 1.

\begin{table}[t]
  \centering
  \caption{Summary of the soft Wilson lines shown in
    Fig.~\ref{figwilson}.\label{table1}}
  \begin{tabular}{ccc} \\ \hline
type & Wilson line& Fourier transform \\ \hline \hline
(a): & $\displaystyle Y=\exp[\frac{1}{n\cdot \mathcal{P}
  +i\epsilon} (-g n\cdot A_s ) \Bigr]$ & $ \displaystyle Y(x)=P\exp
\Bigl[ ig\int_{-\infty}^x ds n\cdot A_s (ns) \Bigr]$\\
(b): & $\displaystyle \tilde{Y}^{\dagger}=\exp \Bigl[ -gn\cdot
A_s \frac{1}{n\cdot \mathcal{P}^{\dagger} +i\epsilon} \Bigr]$&
$\displaystyle \tilde{Y}^{\dagger} (x) =P\exp \Bigl[ig\int_x^{\infty}
ds n\cdot A_s (ns) \Bigr]$ \\
(c):& $\displaystyle \tilde{Y} = \exp \Bigl[ \frac{1}{n\cdot
  \mathcal{P} -i\epsilon} (-gn\cdot A_s)\Bigr]$ & $\displaystyle
\tilde{Y} (x) = \overline{P} \exp \Bigl[-ig\int_x^{\infty} ds n\cdot
A_s (ns) \Bigr]$\\
(d):& $\displaystyle Y^{\dagger} = \exp \Bigl[-g n\cdot A_s
\frac{1}{n\cdot \mathcal{P}^{\dagger} -i\epsilon} \Bigr]$ &
$\displaystyle Y^{\dagger} (x) = \overline{P} \exp \Bigl[
-ig\int_{-\infty}^x ds n\cdot A_s (ns) \Bigr]$ \\ \hline
  \end{tabular}
  \end{table}

The factorization of the usoft interaction in
$\mathrm{SCET}_{\mathrm{I}}$ can be achieved by redefining the
collinear fields as
\begin{equation}
  \xi \rightarrow Y (\tilde{Y}) \xi, \ \
\overline{\xi} \rightarrow \overline{\xi}Y^{\dagger}
(\tilde{Y}^{\dagger}), \ \ \chi \rightarrow \overline{Y}
(\tilde{\overline{Y}} ) \chi, \ \overline{\chi} \rightarrow
\overline{\chi} \overline{Y}^{\dagger}
(\tilde{\overline{Y}}^{\dagger}),
\end{equation}
where the choice of the Wilson lines depends on the physical processes
under consideration. In order to choose appropriate soft Wilson lines,
we should consider whether the collinear particle, which emits soft
gluons, is a particle or an antiparticle, and where it is
directed. The possible soft Wilson lines listed in Table~\ref{table1}
are the building blocks to construct soft Wilson lines in physical
processes. The $i\epsilon$ prescription is irrelevant
if we consider only tree-level processes, but it is critical when the
radiative corrections are investigated to see the scaling behavior of
the soft Wilson lines.

\section{Derivation of the soft Wilson operators in SCET \label{sec3}}
Now that the analytic structure of the soft Wilson lines is
completely known, we can investigate which paths the soft  Wilson lines
take in various physical processes. We consider deep inelastic
scattering, the Drell-Yan process, the jet production in $e^+ e^-$
collisions, and the $\pi$-$\gamma$ form factor near the
phase boundary, in which the same type of the soft Wilson line
appears, but with a different analytic structure. In all these
processes, we put all the collinear particles on the light cone,
namely either in the $n^{\mu}$ or $\overline{n}^{\mu}$ direction and
consider the leading contributions in SCET.
The effect of the soft gluon emission can be expressed as the matrix
elements of the soft Wilson-line operator along the specific path
which is determined by the kinematics. By applying the two-step
matching in SCET near the phase space boundaries, we derive the soft
Wilson-line operator in $\mathrm{SCET}_{\mathrm{II}}$. We choose the
collinear particles in each process in such a way that the resultant
soft Wilson lines have the same form in Eq.~(\ref{keta}).

We consider the electromagnetic current $j^{\mu}=\overline{\psi}
\gamma^{\mu} \psi$ for these processes, which can be transparently
extended to other currents. The first step is to express
the current in $\mathrm{SCET}_{\mathrm{I}}$ by matching the full
theory onto $\mathrm{SCET}_{\mathrm{I}}$. The fermion field $\psi$ is
replaced by the corresponding collinear fermions $\xi$ ($\chi$) in the
$n^{\mu}$ ($\overline{n}^{\mu}$) direction with the appropriate Wilson
coefficient. Before we go down to
$\mathrm{SCET}_{\mathrm{II}}$ below the scale $p_X^2 \sim Q^2 (1-x)$,
we factorize the usoft interactions by redefining the collinear
fields. The form of the usoft Wilson lines depends on the physical
processes under consideration. After the redefinition, we scale down
to $\mathrm{SCET}_{\mathrm{II}}$ and extract the contribution of the
soft gluon emission.

\subsection{Deep inelastic scattering}

In deep inelastic scattering, we choose the Breit frame in which the
proton moves in the $\overline{n}^{\mu}$ direction and the final state
consists of a jet in the $n^{\mu}$ direction. The momentum transfer to
the hadronic system is $Q^2 =-q^2$ which is the large scale, and
$q^{\mu} =(\overline{n}\cdot q, q_{\perp}, n\cdot q) =(Q,0,-Q)$. The
Bjorken variable $x$ is defined as $ x= Q^2/(2p\cdot q) \approx
-n\cdot q/n\cdot p\sim Q/n\cdot p$, where $p^{\mu}$ is the momentum of
the proton. We consider the region with $x\sim 1$, where almost all the
momentum of the proton is carried by a parton undergoing a hard
process. In this case, the final-state particles have invariant mass
$p_X^2 =(p+q)^2 \sim Q^2 (1-x)$. The momenta $p^{\mu}$, and
$p_X^{\mu}$ in the Breit frame scale as
\begin{equation}
p^{\mu}=(\overline{n}\cdot p, p_{\perp}, n\cdot p) \sim
\Bigl(\frac{\Lambda^2}{Q}, \Lambda, \frac{Q}{x} \Bigr),\  \ p_X^{\mu}
\sim (Q, \Lambda, Q\frac{1-x}{x}).
\end{equation}
We integrate out the degrees of freedom of order
$Q^2$ to obtain $\mathrm{SCET}_{\mathrm{I}}$, and subsequently we
integrate out the degrees of freedom of order $Q^2 (1-x)$ to go down
to $\mathrm{SCET}_{\mathrm{II}}$.

\begin{figure}[b]
\begin{center}
\epsfig{file=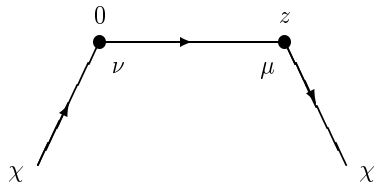, width=5.0cm}
\end{center}
\vspace{-0.6cm}
\caption{Forward Compton scattering amplitude in deep inelastic
  scattering. At the spacetime point 0, a quark $\chi$ from
  $-\infty$ is annihilated and a quark $\xi$ is produced and moves to
  $z$. At $z$, a quark $\xi$ is annihilated, and a quark $\chi$ is
  produced and moves to $\infty$.} 
\label{figfordis}
\end{figure}

The hadronic tensor in the full theory is defined as
\begin{equation}
  W^{\mu \nu} (p,q)= \frac{1}{\pi} \mathrm{Im}\, T^{\mu\nu} (p,q),
\end{equation}
and $T^{\mu\nu}$ is the
spin-averaged matrix element of the forward Compton scattering
amplitude which is given by
\begin{equation}
  T^{\mu\nu} (p,q) = \langle p| \hat{T}^{\mu\nu} (q)
  |p\rangle_{\mathrm{spin\ av.}}, \ \ \hat{T}^{\mu\nu}= i \int d^4 z
  e^{iq\cdot z} T[j^{\mu} (z) j^{\nu}   (0)],
\end{equation}
where $j^{\mu} (x) =\overline{\psi} \gamma^{\mu} \psi (x)$ is the
electromagnetic current for electroproduction.
If the virtuality of the intermediate
states is of order $Q^2$, we can perform the conventional
operator product expansion in which the product of the current
operators is expanded in terms of local operators in powers of $1/Q$
\cite{Bauer:2002nz}. However, when the virtuality of the final-state
particles is of order $p_X^2 \approx Q^2
(1-x) \sim Q\Lambda$, the operator product expansion breaks
down. Still, we can apply SCET and express $\hat{T}^{\mu\nu}$ in terms
of the bilocal operators \cite{Manohar:2003vb}. In SCET, the
momenta of the external collinear particles scale as
\begin{eqnarray}
p_{\bar{n}}^{\mu} &=& (\overline{n}\cdot p_{\bar{n}},
 p_{\bar{n}\perp}, n\cdot p_{\bar{n}}) \sim
 (\frac{\Lambda^2}{Q},\Lambda, \frac{Q}{x}), \nonumber \\
p_n^{\mu} &=& (\overline{n} \cdot p_n, p_{n\perp}, n\cdot p_n) \sim
 (Q, \Lambda, \frac{1-x}{x}Q).
\end{eqnarray}

\begin{figure}[t]
\begin{center}
\epsfig{file=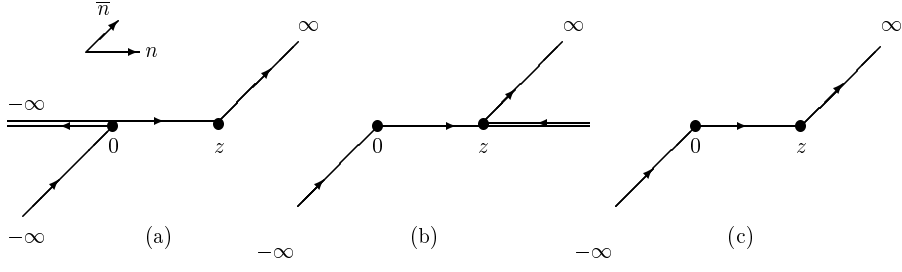, width=12.0cm}
\end{center}
\vspace{-0.6cm}
\caption{The description of the usoft Wilson lines in deep inelastic
  scattering. The usoft Wilson lines with $\xi$ (a) going to
  $-\infty$, (b) going to $\infty$,   (c) the resultant usoft Wilson
  lines from (a) and (b).} 
\label{figdis}
\end{figure}

The time-ordered product of $\hat{T}^{\mu\nu}$ is schematically shown
in Fig.~\ref{figfordis}. A quark $\chi$ from $-\infty$ is
annihilated at point 0, which is described by the current
$\overline{\xi} W \gamma^{\nu} \overline{W}^{\dagger} \chi $. Here $W$
and $\overline{W}$ are the collinear Wilson lines which are defined as 
\begin{equation}
W=\sum_{\mathrm{perm}} \exp \Bigl[ \frac{1}{\overline{n}\cdot
  \mathcal{P}} (-g\overline{n} \cdot A_n) \Bigr], \
  \overline{W}=\sum_{\mathrm{perm}}  \exp \Bigl[ \frac{1}{n\cdot
  \mathcal{P}} (-gn \cdot A_{\bar{n}}) \Bigr],
\end{equation}
where $\mathcal{P}^{\mu}$ is the label momentum operator.  On the
other hand, at point $z$, a quark $\chi$ is produced moving to
$\infty$, which is described by the current $\overline{\chi}
\overline{W} \gamma^{\mu} W^{\dagger} \xi$. The usoft Wilson lines
associated with the collinear field $\chi$ are uniquely determined by
the behavior of the external states. However, the prescription of
the usoft Wilson line associated with the quark $\xi$ is subtle
because the quark $\xi$ is a particle in the intermediate state in the
forward Compton scattering amplitude. 

Physically, as shown in Fig.~\ref{figfordis}, usoft gluons are
emitted from the collinear quarks, and the usoft gluons in the
$n^{\mu}$ direction are emitted from the quark $\xi$ between 0 and
$z$. In order to achieve this requirement, there are two possibilities
which are shown in Fig.~\ref{figdis} (a) and (b), in which the soft
Wilson lines are represented in spacetime. The lines with arrows
represent the soft Wilson lines produced by the collinear particles
moving in the specified direction. In Fig.~\ref{figdis} (a),  a quark
$\xi$ is produced at point 0, moves to $-\infty$. And it moves back
from $-\infty$ to point $z$ and is annihilated. A
second possibility is shown in Fig.~\ref{figdis} (b) in which  a quark
$\xi$ is produced at point 0, moves to $\infty$.  And it moves from
$\infty$ back to $z$ to be annihilated. In the
region where two soft Wilson lines overlap [($-\infty$, 0) in 
Fig.~\ref{figdis} (a), and ($z$, $\infty$) in
Fig.~\ref{figdis} (b)], the soft Wilson lines cancel and the
corresponding Wilson lines in both cases are equivalent to the soft
Wilson lines in Fig.~\ref{figdis} (c). Therefore for the collinear
fermions in the intermediate state, we can factorize the usoft
interactions using either $Y$ or $\tilde{Y}$, and it does not affect
the physical properties such as the radiative corrections. 
As we will see later in computing the radiative corrections, when the
separation between 0 and $z$ is lightlike in the $n^{\mu}$ direction,
the radiative corrections depend only on the analytic structure of the
soft Wilson lines in the $\overline{n}^{\mu}$ direction, and
independent of the analytic structure of the soft Wilson lines in the
$n^{\mu}$ direction. The above statement holds true when $n^{\mu}$ and
$\overline{n}^{\mu}$ are switched. 

From the prescription described above, we factorize the usoft
interactions by redefining the collinear fields as
\begin{eqnarray}
    \label{eq:redef}
\overline{\xi} W \gamma^{\nu} \overline{W}^{\dagger} \chi: &&
    \overline{\xi} \rightarrow \overline{\xi}Y^{\dagger}
    (\tilde{Y}^{\dagger}), \ 
    \ A_n^{\mu}
    \rightarrow Y (\tilde{Y}) A_n^{\mu} Y^{\dagger}
    (\tilde{Y}^{\dagger}),  \  \chi 
    \rightarrow \overline{Y} \chi, \ \
    A_{\overline{n}}^{\mu} \rightarrow \overline{Y}
    A_{\overline{n}}^{\mu} \overline{Y}^{\dagger}, \nonumber \\
\overline{\chi}
\overline{W} \gamma^{\mu} W^{\dagger} \xi: && \overline{\chi}
    \rightarrow \overline{\chi}
    \tilde{\overline{Y}}^{\dagger}, \ A_{\overline{n}}^{\mu}
    \rightarrow \tilde{\overline{Y}} A_{\overline{n}}^{\mu}
    \tilde{\overline{Y}}^{\dagger}, \ \xi \rightarrow Y (\tilde{Y})
    \xi, \     A_n^{\mu} \rightarrow Y (\tilde{Y}) A_n^{\mu}
    Y^{\dagger} (\tilde{Y}^{\dagger}).
\end{eqnarray}
For the collinear particles in the $n^{\mu}$ direction, the soft
Wilson lines without (with) a parenthesis corresponds to the
prescription described by Fig.~\ref{figdis} (a) [Fig.~\ref{figdis}
  (b)]. From now on, we describe the soft Wilson lines shown in
Fig.~\ref{figdis} (a). The current operator in
$\mathrm{SCET}_{\mathrm{I}}$ after the usoft factorization is written
as 
\begin{eqnarray} \label{j1}
  j^{\mu} (z) &=& C(Q)  \Bigl[ e^{i(\overline{n}\cdot p_n n\cdot z/2
-n\cdot p_{\bar{n}} \overline{n}\cdot z/2  )}
 \overline{\xi} W Y^{\dagger} \gamma^{\mu}  \overline{Y}
 \overline{W}^{\dagger}   \chi (z) \nonumber \\
&&+  e^{i(-\overline{n}\cdot p_n n\cdot z/2
+n\cdot p_{\bar{n}} \overline{n}\cdot z/2 )}
\overline{\chi} \overline{W}
 \tilde{\overline{Y}}^{\dagger}  \gamma^{\mu} Y W^{\dagger}
 \xi (z) \Bigr],
\end{eqnarray}
where the exponential factors are the label momenta of order $Q$. The
Wilson coefficient $C(Q)$ is obtained from  the matching of the full
theory onto $\mathrm{SCET}_{\mathrm{I}}$, and is given by
\cite{Manohar:2003vb} 
\begin{equation}
C(Q) = 1+\frac{\alpha_s (Q) C_F}{4\pi} \Bigl( -8 +\frac{\pi^2}{6}
\Bigr).
\end{equation}

The time-ordered product of the two currents in
$\mathrm{SCET}_{\mathrm{I}}$  is given by
\begin{eqnarray} \label{tmunu1}
  \hat{T}^{\mu\nu}_{\mathrm{I}} &=&i\int d^4 z e^{iq\cdot z} T[j^{\mu}
  (z) j^{\nu} (0)]   \nonumber \\
&=& i C^2 (Q) \int d^4 z  e^{i(n\cdot q + n\cdot p_{\bar{n}})
  \overline{n} \cdot   z/2} e^{i(\overline{n}\cdot q
 -\overline{n} \cdot p_n) n\cdot z/2}  \nonumber \\
&&\times T\Bigl[ \Bigl(\overline{\chi} \overline{W}
\tilde{\overline{Y}}^{\dagger}  \gamma^{\mu}  Y W^{\dagger}
\xi\Bigr)(z) \Bigl(  \overline{\xi} W Y^{\dagger} \gamma^{\nu}
\overline{Y} \overline{W}^{\dagger} \chi\Bigr) (0) \Bigr]
  \nonumber   \\
&=& i C^2 (Q)\int d^4 z e^{iQ(1-x) \overline{n} \cdot z/2} T \Bigl[
\Bigl(\overline{\chi} \overline{W} \tilde{\overline{Y}}^{\dagger}
\gamma^{\mu} YW^{\dagger} \xi\Bigr)(z) \Bigl( \overline{\xi} W
Y^{\dagger}  \gamma^{\nu} \overline{Y} \overline{W}^{\dagger}
  \chi\Bigr) (0) \Bigr].
\end{eqnarray}
Here we use the fact that, for label momenta $p$, $p^{\prime}$, and
residual momenta $k$, $k^{\prime}$,
\begin{equation}
\int d^4 z e^{i(p-p^{\prime} +k-k^{\prime})} =\delta_{p,p^{\prime}}
\int d^4 z e^{i(k-k^{\prime}) z}.
\end{equation}
The second exponential factor in the second line of Eq.~(\ref{tmunu1})
is converted to $\delta_{\bar{n}\cdot q, \bar{n}\cdot p_n} =\delta
_{Q,\bar{n}\cdot p_n}$, but the first exponential factor needs a
careful treatment. Since $n\cdot q +n\cdot p_{\bar{n}} = -Q+Q/x = -Q
+(Q+Q(1-x)/x)$, the label momentum of order $Q$ turns into a Kronecker
delta, and there is a slight mismatch in the large momenta in such a
way that the difference of the two large momenta produces a small
momentum of order $Q(1-x)$ for $x\sim 1$ and we explicitly keep it in
Eq.~(\ref{tmunu1}).

Since there are no collinear particles in the $n^{\mu}$ direction in
the proton, we can write \cite{Bauer:2001yt}
\begin{equation}  \label{eq:jk}
  \langle 0| T [W^{\dagger} \xi] (z) [\overline{\xi} W](0)|0\rangle
  \equiv   i\int \frac{d^4 k}{(2\pi)^4} e^{-ik\cdot z} J_P(k)
  \frac{\FMslash{n}}{2},
\end{equation}
which defines the jet function $J_P (k)$. Here the label $P$ is the
sum of the label momenta carried by the collinear fields. This is
obtained by integrating out the degrees of freedom with $p_X^2 \sim
Q^2 (1-x)$.

In $\mathrm{SCET}_{\mathrm{II}}$ after integrating out the degrees of
freedom of order $Q^2 (1-x)$ and renaming the usoft Wilson line $Y$
as the soft Wilson line $S$, the time-ordered product in
$\mathrm{SCET}_{\mathrm{II}}$ is written as
\begin{eqnarray} \label{tmn2}
  \hat{T}^{\mu\nu}_{\mathrm{II}} &=& -\int d\omega C^2 (\omega) \int
  \frac{d^4 k}{(2\pi)^4} \int d^4 z e^{i\Bigl( Q(1-x) -n\cdot k\Bigr)
  \overline{n}\cdot z/2} e^{-ik_{\perp} \cdot z_{\perp} -i\bar{n}\cdot
  k n\cdot z/2} \nonumber \\
&&\times J_P (n\cdot k) T \Bigl[
\Bigl(\overline{\chi} \overline{W} \tilde{\overline{S}}^{\dagger}
\gamma^{\mu} S \Bigr)(z) \delta (\omega - \mathcal{P}_+ )
  \frac{\FMslash{n}}{2} \Bigl(
S^{\dagger}  \gamma^{\nu} \overline{S} \overline{W}^{\dagger}
  \chi\Bigr) (0) \Bigr],
\end{eqnarray}
where $\mathcal{P}_+ = n\cdot \mathcal{P} +n\cdot
\mathcal{P}^{\dagger}$. Using the fact that
\begin{equation}
  \int \frac{d^4 k d^4 z}{(2\pi)^4} e^{-ik_{\perp} \cdot z_{\perp}
  -i\bar{n}\cdot   k n\cdot z/2} = \frac{1}{2\pi} \int dn\cdot k  \,
  d^4 z   \delta^2 (z_{\perp}) \delta (n\cdot z/2) =\frac{1}{4\pi} \int
  dn\cdot k  \,  d\overline{n}\cdot z,
\end{equation}
Eq.~(\ref{tmn2}) can be written as
\begin{eqnarray} \label{tmnf}
\hat{T}^{\mu\nu}_{\mathrm{II}} &=& -\int d\omega C^2 (\omega)  \int
\frac{dn\cdot k d\overline{n}\cdot z}{4\pi}  e^{i\Bigl(
  Q(1-x) -n\cdot k  \Bigr)   \overline{n}\cdot z/2} J_P (n\cdot
k)  \nonumber \\
&&\times T \Bigl[
\Bigl(\overline{\chi} \overline{W} \tilde{\overline{S}}^{\dagger}
\gamma^{\mu} S \Bigr)(\overline{n}\cdot z) \delta (\omega -
\mathcal{P}_+ ) \frac{\FMslash{n}}{2}  \Bigl( S^{\dagger}
\gamma^{\nu} \overline{S} \overline{W}^{\dagger}  \chi\Bigr) (0)
\Bigr] \nonumber \\ 
&=& -\int d\omega C^2 (\omega)  \int
\frac{dn\cdot k d\overline{n}\cdot z}{4\pi} \int d\eta e^{i\Bigl(
  Q(1-x) -n\cdot k -\eta   \Bigr)   \overline{n}\cdot z/2} J_P (n\cdot
k)  \nonumber \\
&&\times T \Bigl[
\Bigl(\overline{\chi} \overline{W} \tilde{\overline{S}}^{\dagger}
\gamma^{\mu} S \Bigr)(0) \delta (\omega - \mathcal{P}_+ )
\frac{\FMslash{n}}{2}  \delta (\eta
+n\cdot i\partial) \Bigl(
S^{\dagger}  \gamma^{\nu} \overline{S} \overline{W}^{\dagger}
  \chi\Bigr) (0) \Bigr] \nonumber \\
&=& -\int d\omega  C^2(\omega) \int dn\cdot k \int d\eta  J_P (n\cdot
k) \delta \Bigl(Q(1-x) -n\cdot k -\eta \Bigr) \nonumber \\
&&\times  T\Bigl[\Bigl( \overline{\chi} \overline{W} \Bigr)_{\alpha}
\delta (\omega - \mathcal{P}_+ ) \gamma^{\mu} \frac{\FMslash{n}}{2}
\gamma^{\nu}  \Bigl(\overline{W}^{\dagger}
    \chi\Bigr)_{\beta}  \Bigl( \tilde{\overline{S}}^{\dagger} S \delta
    (\eta+n\cdot  i\partial) S^{\dagger} \overline{S}
    \Bigr)_{\alpha\beta}\Bigr] \nonumber \\
&\rightarrow& -\int d\omega C^2(\omega)
T\Bigl[ \Bigl( \overline{\chi} \overline{W} \Bigr)
    \delta (\omega - \mathcal{P}_+ )
    \gamma^{\mu} \frac{\FMslash{n}}{2}  \gamma^{\nu}
    \Bigl(\overline{W}^{\dagger}     \chi\Bigr) (0) \nonumber \\
&&\times \int dn\cdot k J_P (n\cdot k)  K\Bigl( Q(1-x) -n\cdot
k\Bigr) \Bigr],
\end{eqnarray}
where $K(\eta)$ is the soft Wilson-line operator which is defined as
\begin{eqnarray}
  K(\eta) &=& \frac{1}{N} \mathrm{tr} \Bigl(
    \tilde{\overline{S}}^{\dagger} S \delta  (\eta+n\cdot  i\partial)
    S^{\dagger}     \overline{S}   \Bigr) \nonumber \\
&=& \frac{1}{N} \mathrm{tr} \exp \Bigl[ -g\overline{n} \cdot A_s
    \frac{1}{\overline{n} \cdot \mathcal{P}^{\dagger} +i\epsilon}
    \Bigr] \exp \Bigl[\frac{1}{n\cdot \mathcal{P} +i\epsilon} (-g
    n\cdot A_s) \Bigr] \delta (\eta +n\cdot i\partial) \nonumber \\
&&\times \exp \Bigl[-gn\cdot A_s \frac{1}{n\cdot \mathcal{P}^{\dagger}
    -i\epsilon} \Bigr] \exp \Bigl[\frac{1}{\overline{n} \cdot
    \mathcal{P} +i\epsilon} (-g \overline{n} \cdot A_s) \Bigr],
\end{eqnarray}
while $K(\eta)$ becomes
\begin{eqnarray}
  K(\eta) &=& \frac{1}{N} \mathrm{tr} \Bigl(
    \tilde{\overline{S}}^{\dagger} \tilde{S} \delta  (\eta+n\cdot
    i\partial) \tilde{S}^{\dagger}     \overline{S}   \Bigr) \nonumber
    \\ 
&=& \frac{1}{N} \mathrm{tr} \exp \Bigl[ -g\overline{n} \cdot A_s
    \frac{1}{\overline{n} \cdot \mathcal{P}^{\dagger} +i\epsilon}
    \Bigr] \exp \Bigl[\frac{1}{n\cdot \mathcal{P} -i\epsilon} (-g
    n\cdot A_s) \Bigr] \delta (\eta +n\cdot i\partial) \nonumber \\
&&\times \exp \Bigl[-gn\cdot A_s \frac{1}{n\cdot \mathcal{P}^{\dagger}
    +i\epsilon} \Bigr] \exp \Bigl[\frac{1}{\overline{n} \cdot
    \mathcal{P} +i\epsilon} (-g \overline{n} \cdot A_s) \Bigr],
\end{eqnarray}
if we use the prescription described by Fig.~\ref{figdis} (b). 

In Eq.~(\ref{tmnf}), the third equality is obtained because the soft
particles are  decoupled from the collinear particles and $\alpha$,
$\beta$ are color indices. And in the final result we take the color
average of the soft Wilson-line operator, represented by the trace
since the soft Wilson line will be sandwiched between the vacuum
states. Care must be taken in separating the soft part from the
collinear part in Eq.~(\ref{tmnf}) since the operator $\delta (\eta
+n\cdot i\partial)$ acts on $S^{\dagger} \overline{S}
\overline{W}^{\dagger} \chi$. When $n\cdot i\partial$ acts on
$\overline{W}^{\dagger} \chi$, it produces the momentum of order
$\Lambda$, that is, $n\cdot i\partial \overline{W}^{\dagger} \chi \sim
\Lambda \overline{W}^{\dagger}\chi$ since the largest momentum of order
$Q$ is already extracted as the label momentum and the remaining
momentum can be of order $\Lambda$. However, due to the
reparameterization invariance \cite{Chay:2002vy,Manohar:2002fd}, we
can transform away the 
momentum component of order $\Lambda$ in the collinear fields
$\overline{W}^{\dagger} \chi$. Therefore the derivative $n\cdot
i\partial$ applied to the collinear fields can be neglected.
As a result, the time-ordered product $\hat{T}^{\mu\nu}$ is given
as the convolution of the Wilson coefficient $C(\omega)$, the collinear
operator, the jet function which can be obtained by matching between
$\mathrm{SCET}_{\mathrm{I}}$ and $\mathrm{SCET}_{\mathrm{II}}$, and
finally the soft Wilson-line operator.

One important comment is in order about the relation between the
forward scattering amplitude and the scattering cross sections. In our
approach in which we consider the forward scattering amplitude, we
take the discontinuity of the hadronic tensor $T^{\mu\nu}$ in the
final step, which is given by
\begin{eqnarray} \label{disc}
  2 \, \mathrm{Im}\, T^{\mu\nu} &=& \sum_f \int d\Pi_f \langle p|
  J^{\mu}   (-q) |f\rangle \langle f|J^{\nu} (q)
  |p\rangle_{\mathrm{spin \ av.}}   \nonumber \\
&=& \sum_f \int d\Pi_f \langle p| \overline{T} \Bigl[ \overline{\chi}
  \overline{W} \overline{Y}^{\dagger}  \gamma^{\mu}
  \tilde{Y}W^{\dagger} \xi \Bigr] |f\rangle \langle f| T
  \Bigl[ \overline{\xi} W \tilde{Y}^{\dagger}  \gamma^{\nu}
 \overline{Y} \overline{W}^{\dagger}
  \chi\Bigr]|p\rangle_{\mathrm{spin \ av.}},
\end{eqnarray}
where the first matrix element $\langle p| \overline{T} \Bigl[
\overline{\chi} \overline{W} \overline{Y}^{\dagger}  \gamma^{\mu}
\tilde{Y}W^{\dagger} \xi \Bigr] |f\rangle$ is the hermitian
conjugate of $\langle f| T
\Bigl[ \overline{\xi} W \tilde{Y}^{\dagger}  \gamma^{\nu}
\overline{Y} \overline{W}^{\dagger}
\chi\Bigr]|p\rangle$, in which the time ordering
becomes the anti-time ordering. We may develop a calculational
technique to compute the radiative corrections for the matrix
elements squared \cite{Korchemsky:1992xv,collins82}, but it is more
convenient to use the optical theorem. That is, we consider the
time-ordered products $\hat{T}^{\mu\nu}$, and we can compute the
radiative corrections using the conventional Feynman rules. At the end
we take the discontinuity of $\hat{T}^{\mu\nu}$. 

This approach is also advantageous if we use
SCET. In SCET, the hadronic tensor is described by the gauge-invariant
operators, whose matrix elements can be attributed to the
nonperturbative effects. We first match the current between the full
theory and $\mathrm{SCET}_{\mathrm{I}}$ near $\mu^2 = Q^2$, then we
integrate out the degrees of freedom of order $\mu^2 \sim Q^2 (1-x)$
to obtain nonlocal, but gauge-invariant operators in
$\mathrm{SCET}_{\mathrm{II}}$. In this procedure, the jet function
$J_P (k)$ is obtained to a desired order in $\alpha_s$, but all the
radiative corrections for $J_P (k)$ come from collinear interactions,
not from usoft interactions, otherwise the radiative corrections
cannot be integrated out. Therefore when we take the imaginary part of
$T^{\mu\nu}$, the
contribution only comes from the discontinuity of the jet function
$J_P (k)$. Then  we consider the contributions from soft
particles in $\mathrm{SCET}_{\mathrm{II}}$, which are decoupled from
the collinear sector. The soft
radiative corrections do not have discontinuity at least at one loop
order. To summarize, the strategy for computing radiative soft
interactions is to compute the radiative corrections in the left-hand
side of Eq.~(\ref{disc}), in which all the operators are time-ordered,
instead of computing the right-hand side of Eq.~(\ref{disc}). Finally
we take its discontinuity to obtain the hadronic tensor.

This distinction has also been realized in the consideration using the
full theory. For example, in Ref.~\cite{Korchemskaya:1992je}, the
radiative correction is computed using the time-ordered products. On
the other hand, in Ref.~\cite{Korchemsky:1992xv}, it is computed using
the matrix element squared and in obtaining the final result, the
authors carefully distinguished the analytic structure of the soft
Wilson lines by taking a hermitian conjugate from the results in
Ref.~\cite{Korchemskaya:1992je} for the part on the opposite side
of the physical cut. We would like to stress that both
approaches are equivalent and the same results should be
obtained. However, we choose the approach with the forward scattering
amplitude in which only time-ordered products appear and we can
compute radiative corrections using the Feynman rules derived from the
form $K(\eta)$. 

\subsection{Drell-Yan processes}
In Drell-Yan processes $p\overline{p} \rightarrow \ell \overline{\ell}
+X$, lepton pairs, preferably muon pairs, are
produced in proton-antiproton scattering. This process can be regarded
as $p\overline{p} \rightarrow \gamma^* X$, where $\gamma^*$ is a
virtual photon which eventually produces a lepton pair. Let the
momenta of the proton (antiproton) be $p$ ($\overline{p}$) in the
$\overline{n}^{\mu}$ ($n^{\mu}$) direction. The kinematic variables in
this process are given by
\begin{eqnarray}
s &=& (p+\overline{p})^2 = n\cdot p \overline{n}\cdot \overline{p}, \
Q^2 = q^2, \ \tau = \frac{Q^2}{s}, \nonumber \\
x&=&\frac{Q^2}{2p\cdot q} =\frac{Q^2}{n \cdot p \overline{n}\cdot q} =
\frac{n \cdot q}{n \cdot p}, \ \overline{x}
=\frac{Q^2}{2\overline{p} \cdot q} = \frac{\overline{n}\cdot
  q}{\overline{n}\cdot \overline{p}},
\end{eqnarray}
where $q^{\mu} =(\overline{n}\cdot q, q_{\perp}, n\cdot q) =(Q,0,Q)$
is the momentum of the virtual photon.  The momentum
$p_X=p+\overline{p}-q$ of the final hadronic system satisfies the
relation
\begin{equation}
p_X^2 = s+Q^2 -2p\cdot q-2\overline{p} \cdot q = Q^2
\Bigl(1+\frac{1}{\tau} -\frac{1}{x} -\frac{1}{\overline{x}} \Bigr).
\end{equation}
If one of the partons approaches the endpoint, say, $x \sim
1$, ($\overline{x}$ away from the endpoint) $\tau = x\overline{x}
\approx x$, and $p_X^2 \approx Q^2
(1-x)(1-\overline{x})/\overline{x}$. The momenta scale as
\begin{eqnarray}
p^{\mu} &=& p_{\bar{n}}^{\mu} = (\overline{n}\cdot p, p_{\perp},
n\cdot p) \sim \Bigl(\frac{\Lambda^2}{Q},\Lambda,\frac{Q}{x} \Bigr),
\nonumber \\ 
\overline{p}^{\mu} &=&   (\overline{n}\cdot \overline{p},
\overline{p}_{\perp}, n\cdot \overline{p}) \sim
\Bigl(\frac{Q}{\overline{x}}, \Lambda, \frac{\Lambda^2}{Q}\Bigr),
\ p_X^{\mu} \sim \Bigl( \frac{1-\overline{x}}{\overline{x}} Q,
\Lambda, (1-x) Q\Bigr).
\end{eqnarray}
Therefore near the endpoint $x\sim 1$, the final-state hadrons move in
the $n^{\mu}$ direction, that is, in the direction of the
antiproton, and we can apply SCET to integrate out the degrees of
freedom of order $Q^2$ and $Q^2(1-x)$ successively.

\begin{figure}[t]
\begin{center}
\epsfig{file=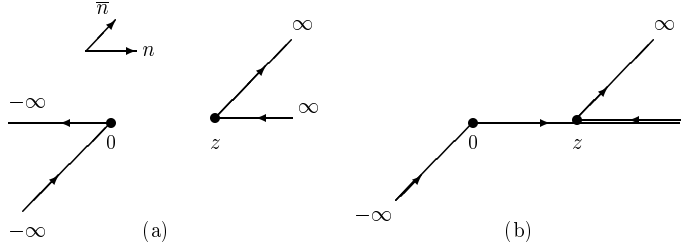, width=9.0cm}
\end{center}
\vspace{-0.6cm}
\caption{(a) The usoft Wilson lines in the Drell-Yan process at lowest
  order. (b) the configuration of the usoft Wilson 
lines for the Drell-Yan process at higher orders in $\alpha_s$, which
is equal to that of deep inelastic scattering, hence giving the same
radiative corrections.} 
\label{figdy}
\end{figure}

The hadronic tensor $W^{\mu\nu}$ is defined in close analogy to deep
inelastic scattering  as
\begin{eqnarray} \label{dyw}
  W^{\mu\nu} &=& \frac{1}{4} \sum_{\mathrm{spins}} \sum_X (2\pi)^4
  \delta^4 (p+\overline{p} -q -p_X) \langle p\overline{p} | j^{\mu}(0)
  |X\rangle \langle X|j^{\nu} (0) |p\overline{p}\rangle \nonumber \\
&=& \frac{1}{4} \sum_{\mathrm{spins}} \int d^4 z e^{-iq\cdot z}
  \langle p\overline{p} |j^{\mu} (z) j^{\nu} (0)|p\overline{p}
  \rangle.
\end{eqnarray}
As in the case of deep inelastic scattering, instead of computing
Eq.~(\ref{dyw}), we consider the forward scattering amplitude which
corresponds to the matrix element of the time-ordered product of the
two currents, which is given as
\begin{equation}
  \mathcal{W} =\int d^4 z e^{-iq\cdot z} T[j^{\mu} (z) j_{\mu} (0)].
\end{equation}
The electromagnetic current in the Drell-Yan process before the usoft
factorization is also given as
\begin{eqnarray} \label{dycur}
  j^{\mu} (z) &=& C(Q)  \Bigl[ e^{-i(\overline{n}\cdot p_n n\cdot z/2
 +n \cdot p_{\bar{n}} \overline{n}\cdot z/2)}
 \overline{\xi} W \gamma^{\mu}  \overline{W}^{\dagger}
  \chi (z) \nonumber \\
&&+ e^{i(\overline{n}\cdot p_n n\cdot z/2 +n \cdot p_{\bar{n}}
 \overline{n}\cdot z/2)}\overline{\chi} \overline{W} \gamma^{\mu}
 W^{\dagger}  \xi (z) \Bigr],
\end{eqnarray}
but the redefinition of the collinear fields to factorize the usoft
interactions becomes different compared to deep inelastic
scattering. In order to compute the forward scattering amplitude in
the Drell-Yan process, the current $ \overline{\xi} W \gamma^{\mu}
\overline{W}^{\dagger}
\chi (z)$ describes a quark $\chi$ and an antiquark $\xi$ coming from
$-\infty$, which are annihilated at $z$. The second current in
Eq.~(\ref{dycur}) describes a quark 
$\chi$ and an antiquark $\xi$, which are created at $z$ and move to
$\infty$. Since the collinear particles which determine the structure
of the usoft Wilson lines are in the initial and the final states, the
prescription is specified without ambiguity, and the usoft Wilson
lines are described in Fig.~\ref{figdy} (a). Fig.~\ref{figdy} (b) will
be explained at the end of this subsection.

With the prescription shown in Fig.~\ref{figdy} (a), the field
redefinition becomes 
\begin{eqnarray}
 \overline{\xi} W \gamma^{\mu}  \overline{W}^{\dagger}
  \chi : &&  \overline{\xi} \rightarrow \overline{\xi} Y^{\dagger}, \
  A_n^{\mu}   \rightarrow YA_n^{\mu} Y^{\dagger}, \ \chi \rightarrow
  \overline{Y}
  \chi,  \ A_{\bar{n}}^{\mu} \rightarrow \overline{Y} A_{\bar{n}}^{\mu}
  \overline{Y}^{\dagger}, \nonumber \\
\overline{\chi} \overline{W} \gamma^{\mu} W^{\dagger}
 \xi : && \overline{\chi} \rightarrow \overline{\chi}
  \tilde{\overline{Y}}^{\dagger}, \ A_{\bar{n}}^{\mu} \rightarrow
  \tilde{\overline{Y}} A_{\bar{n}}^{\mu}
  \tilde{\overline{Y}}^{\dagger}, \ \xi \rightarrow \tilde{Y} \xi, \
  A_n^{\mu} \rightarrow \tilde{Y} A_n^{\mu} \tilde{Y}^{\dagger}.
\end{eqnarray}
Therefore the current operator after the redefinition becomes
\begin{eqnarray}
  j^{\mu} (z) &=& C(Q) \Bigl[  e^{-i(\overline{n}\cdot p_n n\cdot z/2
 +n \cdot p_{\bar{n}} \overline{n}\cdot z/2)}\overline{\xi} W
 Y^{\dagger} \gamma^{\mu}   \overline{Y} \overline{W}^{\dagger} \chi
 (z) \nonumber \\
&&+ e^{i(\overline{n}\cdot p_n n\cdot z/2
 +n \cdot p_{\bar{n}} \overline{n}\cdot z/2)}\overline{\chi}
  \overline{W} \tilde{\overline{Y}}^{\dagger} \gamma^{\mu} \tilde{Y}
  W^{\dagger} \xi (z) \Bigr].
\end{eqnarray}
The time-ordered product of the two currents $\mathcal{W}$
can be written as
\begin{eqnarray} \label{dytree}
\mathcal{W}&=& C(Q)^2 \int d^4 z e^{i(n\cdot p_{\bar{n}} -n\cdot q)
\overline{n} \cdot z/2} e^{i(\overline{n} \cdot p_n -\overline{n} \cdot
q) n\cdot z/2} \nonumber \\
&&\times T\Bigl[ \overline{\chi} \overline{W}
\tilde{\overline{Y}}^{\dagger} \gamma^{\mu} \tilde{Y} W^{\dagger} \xi
(z) \overline{\xi} W Y^{\dagger} \gamma_{\mu} \overline{Y}
\overline{W}^{\dagger} \chi (0) \Bigr],
\end{eqnarray}
where $p_n^{\mu} = \overline{p}^{\mu} -p_X^{\mu} \sim (Q,\Lambda,
Q(1-x)/x)$ is the net collinear momentum in the $n^{\mu}$ direction.
In order to obtain the scattering cross section, we evaluate the matrix
element of $\mathcal{W}$ between the states with a proton and an
antiproton, which corresponds to the forward scattering amplitude. And
we take the discontinuity of this amplitude. However there is a
delicate problem on how to extract the soft Wilson lines. First at
zeroth order in $\alpha_s$, there are no final-state hadrons
($p_X=0$). There is no intermediate scale which separates
$\mathrm{SCET}_{\mathrm{I}}$ and $\mathrm{SCET}_{\mathrm{II}}$, and
there is no requirement that the separation between 0 and $z$ be
lightlike. From now on, we will only consider the Drell-Yan process with
some hadrons in the final state. 

\begin{figure}[b]
\begin{center}
\epsfig{file=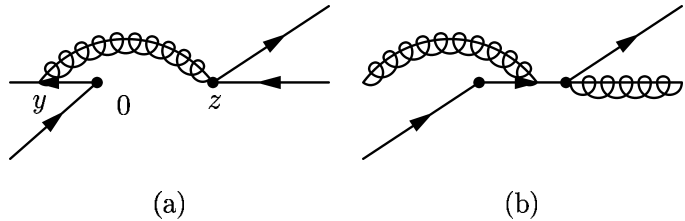, width=9.0cm}
\end{center}
\vspace{-0.6cm}
\caption{The Drell-Yan processes at order $\alpha_s$ in which a
  valence quark comes from the proton and (a) an antiquark, (b) a
  gluon from the antiproton.} 
\label{figdyhi}
\end{figure}

The time-ordered product $\mathcal{W}$ now includes the collinear
Lagrangian $\mathcal{L}_c$ \cite{Bauer:2001yt}, and it can be written
at nontrivial lowest order in $\mathrm{SCET}_{\mathrm{I}}$ as
\begin{eqnarray} \label{dyone}
\mathcal{W}&=& C(Q)^2 \int d^4 z \int d^4 y  e^{i(n\cdot p_{\bar{n}}
  -n\cdot q) \overline{n} \cdot z/2} e^{i(\overline{n} \cdot p_n
  -\overline{n} \cdot q) n\cdot z/2}  \nonumber \\
&&\times T\Bigl[ \overline{\chi} \overline{W} \gamma^{\mu} W^{\dagger}  
\xi (z) \overline{\xi} W \gamma_{\mu} \overline{W}^{\dagger} \chi (0),  
i\mathcal{L}_c (y) \Bigr], 
\end{eqnarray}
and we can include additional $i\mathcal{L}_c$ at higher orders. We
contract the collinear fields in the $n^{\mu}$ directions such that
the resultant operators contains two collinear quarks or gluons in the
$n^{\mu}$ direction. The Drell-Yan processes from Eq.~(\ref{dyone}) at
order $\alpha_s$ are schematically described in Fig.~\ref{figdyhi}. 
In $\mathrm{SCET}_{\mathrm{II}}$, we take the
matrix elements of these operators after Fierz transformation to
parameterize as parton distribution functions in the proton and the
antiproton. Note that we need the gluon operators including
$A_{n\perp}$ to compute the contribution from the gluon distribution
functions in the antiproton, described in Fig.~\ref{figdyhi} (b)
because only the operators with $A_{n\perp}$ contribute. 

The form of the collinear operators is complicated, but there are two
important points in constructing the soft Wilson lines. First, when we
contract the collinear particles, the jet functions are obtained. For
example, when we contract collinear fermions in the $n^{\mu}$
direction, we obtain
\begin{equation}
  \langle 0 |T[ W^{\dagger} \xi (z) \overline{\xi} W (0)]
  |0 \rangle \equiv i\int \frac{d^4 k}{(2\pi)^4}
  e^{-ik\cdot z} \frac{\FMslash{n}}{2} J_P (k),
\end{equation}
and similarly for the collinear gluons. All these jet functions depend
only on $n\cdot k$. Secondly, we have freedom to choose the directions
of the soft Wilson lines associated with the collinear fermions in the
intermediate states, as explained in deep inelastic scattering and we
use this freedom to construct soft Wilson lines. 

Let us take a specific example shown in Fig.~\ref{figdyhi} (a). We can
write the leading collinear Lagrangian as $\mathcal{L}_c (y) =
\overline{\xi} gA_n \cdot \Gamma \FMslash{\overline{n}}\xi (y)$, where
$\Gamma$ is a Dirac structure. We can make it
explicitly gauge invariant, but this form is enough to consider the
soft Wilson lines. The usoft Wilson lines for the external states
$\overline{\chi} (z)$ and $\chi (0)$ are fixed as $\overline{\chi}
(z)\tilde{\overline{Y}}^{\dagger}$ 
and $\overline{Y} \chi (0)$. The intermediate states are obtained by
contracting $\xi (y)$, $\overline{\xi} (0)$, and $A_n (y)$, $A_n
(z)$. We are free to choose $Y$ or $\tilde{Y}$ for the intermediate
states, but once we specify the soft Wilson line for one of the
contracted fields, the usoft Wilson line for the remaining contracted
field should be the same because the contracted fields should start
from the same point. This was also illustrated in Fig.~\ref{figdis}
(a) and (b) for the fermion in the intermediate state. If we choose
the usoft Wilson line for $\xi (z)$ as $\tilde{Y} \xi (z)$, it fixes
all the remaining usoft Wilson lines. Satisfying all these
requirements, $\mathcal{L}_c$ becomes 
\begin{equation}
\Bigl[\overline{\xi} \tilde{Y}^{\dagger} (y) \Bigr] \Bigl[ \tilde{Y}
  gA_n \cdot \Gamma \tilde{Y}^{\dagger} (y) \Bigr]
  \FMslash{\overline{n}} \Bigl[ \tilde{Y} \xi   (y)\Bigr]. 
\end{equation}

The whole set of operators in Eq.~(\ref{dyone}) is written as 
\begin{eqnarray}
\Bigl(\overline{\chi} \overline{W} \tilde{\overline{Y}}^{\dagger} (z) 
\Bigr)  \gamma^{\mu} \Bigl(\tilde{Y} \overbrace{W^{\dagger}
  \tilde{Y}^{\dagger} (z) \Bigr) \Bigl( \tilde{Y} \xi (z) \Bigr)
  \Bigl( \overline{\xi}  gA_n} \cdot \Gamma \Bigr)  
\FMslash{\overline{n}} \underbrace{\xi (y) 
\Bigl( \overline{\xi} W} \tilde{Y}^{\dagger} (0) \Bigr) \gamma_{\mu}
\Bigl(\overline{Y} \overline{W}^{\dagger} \chi (0) \Bigr),  
\end{eqnarray}
where the braces represent the contraction of the corresponding fields
to produce the jet functions. After contracting the fields, if we
express the color structure only, it becomes 
\begin{equation}
\Bigl[\Bigl( \overline{\chi} \overline{W} (z) \Bigr)_{\alpha}
\gamma^{\mu} \Bigl( \tilde{\overline{Y}}^{\dagger} \tilde{Y} (z)
\Bigr)_{\alpha   \beta} (T_a)_{\beta\tau}  \Bigl(\xi (z)
\Bigr)_{\tau} \Bigr] \cdot  \Bigl[ \Bigl(\overline{\xi} (y)
\Bigr)_{\rho} (T_a)_{\rho   \sigma} \Gamma \FMslash{\overline{n}}
\gamma_{\mu} \Bigl( \tilde{Y}^{\dagger} \overline{Y} (0) \Bigr)_{\sigma \nu}
\Bigl(\overline{W}^{\dagger} \chi (0)\Bigr)_{\nu} \Bigr]. 
\end{equation}
Since the collinear quarks $\xi$, $\overline{\xi}$ and $\chi$,
$\overline{\chi}$ should form a color singlet to produce the parton
distribution functions, we project out the color-singlet components
for the usoft part. We can pull out the usoft part since it is
decoupled from the collinear part, and it is written as
\begin{equation} \label{dywil}
\frac{\delta_{\alpha \nu}}{N} \frac{\delta_{\tau\rho}}{N}  \Bigl(
\tilde{\overline{Y}}^{\dagger} \tilde{Y} (\overline{n} \cdot z)
\Bigr)_{\alpha   \beta} (T_a)_{\beta\tau}  (T_a)_{\rho   \sigma}
\Bigl( \tilde{Y}^{\dagger} \overline{Y} (0) \Bigr)_{\sigma \nu} =
\frac{C_F}{N} \frac{1}{N} \mathrm{tr} \, \Bigl(
\tilde{\overline{Y}}^{\dagger} \tilde{Y} (\overline{n} \cdot z) 
\tilde{Y}^{\dagger} \overline{Y} (0) \Bigr), 
\end{equation}
which is described by Fig.~\ref{figdy} (b) with the lightlike
separation between 0 and $z$ because the jet functions depend only on
$n\cdot k$ and all the spacetime points are put on the light cone. 
For Fig.~\ref{figdyhi} (b) where gluons in the antiproton contribute,
the same configuration of the soft Wilson lines is obtained. 
Interestingly enough, Fig.~\ref{figdy} (b) is the prescription 
for deep inelastic scattering, and we expect that the radiative
corrections for deep inelastic scattering and the Drell-Yan process
are the same. If we choose $Y\xi (z)$, $\tilde{Y}$ is replaced by $Y$
in Eq.~(\ref{dywil}), and it yields the same configuration for the
usoft Wilson lines, but corresponds to different a physical process. 

We will not bother to write down $\mathcal{W}$, but there is certainly
a contribution from the soft Wilson lines of the form
\begin{equation}
\int dn\cdot k \tilde{J}_P (n\cdot k) \langle 0|
  T\Bigl[K\Bigl(   Q(1-x) -n\cdot k\Bigr) \Bigr] |0\rangle,
\end{equation}
in the expression of $\mathcal{W}$, where $\tilde{J}_P$ is a collection
of jet functions. And the soft Wilson line operator $K(\eta)$ is given
by 
\begin{eqnarray}
  K(\eta) &=& \frac{1}{N} \mathrm{tr} \,
  \tilde{\overline{S}}^{\dagger} \tilde{S} \delta
  (\eta +n\cdot i\partial) \tilde{S}^{\dagger} \overline{S} \nonumber \\
&=& \frac{1}{N} \mathrm{tr} \, \exp \Bigl[ -g\overline{n}\cdot A_s
  \frac{1}{\overline{n} \cdot \mathcal{P}^{\dagger} +i\epsilon} \Bigr]
  \exp \Bigl[ \frac{1}{n\cdot \mathcal{P} -i\epsilon} (-gn\cdot A_s)
  \Bigr] \nonumber \\
&&\times \delta (\eta +n\cdot i\partial) \exp \Bigl[ -g n\cdot A_s
  \frac{1}{n\cdot \mathcal{P}^{\dagger} +i\epsilon} \Bigr] \exp \Bigl[
  \frac{1}{\overline{n}\cdot \mathcal{P} +i\epsilon} (-g\overline{n}
  \cdot A_s) \Bigr].
\end{eqnarray}

\subsection{Jet production in $e^+ e^-$ scattering}
The soft Wilson lines for the jet production in $e^+ e^- \rightarrow Z
\rightarrow q\overline{q}$ were first derived in
Ref.~\cite{Bauer:2002ie,Bauer:2003di}. Our approach is equivalent,
but different in the fact that we consider the time-ordered products
of the hadronic electromagnetic current to evaluate the forward
scattering amplitude. We choose the frame such that a quark (an
antiquark) is produced in the $n^{\mu}$ ($\overline{n}^{\mu}$)
direction, and the momentum for the final-state particles is given by
$q^{\mu} = (\overline{n}\cdot q, q_{\perp}, n\cdot q) = (Q,0,Q)$. We
consider the case in which a collinear jet in the $\overline{n}^{\mu}$
direction is initiated by the antiquark. If the energy of the
antiquark is close to its maximum, $x=n\cdot \overline{p}/Q \sim 1$,
there are hadrons in the $n^{\mu}$ direction including the quark
jet. The momentum $p_{\bar{n}}^{\mu}$ of the collinear antiquark jet
and the momentum $p_X^{\mu}$ of the hadrons in the $n^{\mu}$ direction
scale as
\begin{equation}
p_{\bar{n}}^{\mu} \sim \Bigl(\frac{\Lambda^2}{xQ}, \Lambda, xQ\Bigr), \
p_X^{\mu} \sim (Q,\Lambda, (1-x)Q).
\end{equation}
Here we can apply the two-step matching of SCET, that is, we first
integrate out the degrees of freedom of order $Q^2$, and go down to
$\mathrm{SCET}_{\mathrm{II}}$ by integrating out the degrees of
freedom of order $Q^2 (1-x)$.

\begin{figure}[t]
\begin{center}
\epsfig{file=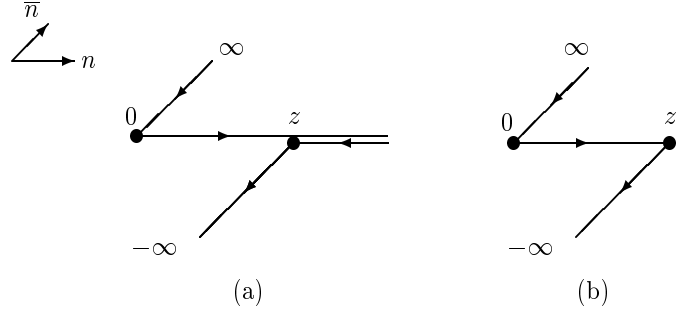, width=9.0cm}
\end{center}
\vspace{-0.6cm}
\caption{(a) The description of the usoft Wilson lines in $e^+e^-$
  collisions. (b) the resultant configuration of the usoft Wilson
  lines.}
\label{figee}
\end{figure}

The hadronic electromagnetic current is given by
\begin{eqnarray}
  j^{\mu} (z) &=& C(Q)  \Bigl[ e^{i(\overline{n}\cdot p_n n\cdot z/2
 +n\cdot p_{\bar{n}} \overline{n}\cdot z/2)}
 \overline{\xi} W \gamma^{\mu}  \overline{W}^{\dagger}
  \chi (z) \nonumber \\
&&+  e^{-i(\overline{n}\cdot p_n n\cdot z/2
 +n\cdot p_{\bar{n}} \overline{n}\cdot z/2)}\overline{\chi}
 \overline{W} \gamma^{\mu} W^{\dagger}  \xi (z) \Bigr].
\end{eqnarray}
In $e^+ e^-$ collisions, the forward scattering amplitude corresponds
to the photon polarization, in which a photon decays into collinear
particles ($\xi$ and $\chi$ at leading order) and they turn into a
photon again. In this case, since both collinear particles are in the
intermediate states, there can arise ambiguities as in the case of
deep inelastic scattering. However, we can think of a collinear jet in
the $\overline{n}^{\mu}$ direction and the remainder becomes hadrons
including the collinear jet in the $n^{\mu}$ direction. Then we can
specify the direction of the antiquark $\chi$, and the description of
the usoft Wilson lines is shown in Fig.~\ref{figee}. Of course, we can
let $\xi$ go to $-\infty$, but as in deep inelastic scattering, that
prescription also gives the usoft Wilson lines in Fig.~\ref{figee}
(b). And we will describe the process using the prescription in
Fig.~\ref{figee} (a). 

The collinear fields are redefined as
 \begin{eqnarray}
\overline{\xi} W \gamma^{\mu}  \overline{W}^{\dagger}  \chi : &&
\overline{\xi} \rightarrow \overline{\xi} \tilde{Y}^{\dagger}, \
A_n^{\mu}   \rightarrow \tilde{Y} A_n^{\mu}  \tilde{Y}^{\dagger}, \
\chi \rightarrow  \tilde{\overline{Y}}  \chi,  \ A_{\bar{n}}^{\mu}
\rightarrow  \tilde{\overline{Y}} A_{\bar{n}}^{\mu}
\tilde{\overline{Y}}^{\dagger}, \nonumber \\
\overline{\chi} \overline{W} \gamma^{\mu} W^{\dagger}
 \xi : && \overline{\chi} \rightarrow \overline{\chi}
 \overline{Y}^{\dagger}, \ A_{\bar{n}}^{\mu} \rightarrow
\overline{Y} A_{\bar{n}}^{\mu} \overline{Y}^{\dagger}, \ \xi
  \rightarrow \tilde{Y} \xi,  \  A_n^{\mu} \rightarrow \tilde{Y}
  A_n^{\mu}   \tilde{Y}^{\dagger},
 \end{eqnarray}
and the current is given by
\begin{eqnarray}
  j^{\mu} (z) &=& C(Q)  \Bigl[ e^{i(\overline{n}\cdot p_n n\cdot z/2
 +n\cdot p_{\bar{n}} \overline{n}\cdot z/2)}
 \overline{\xi} W \tilde{Y}^{\dagger}\gamma^{\mu}
 \tilde{\overline{Y}} \overline{W}^{\dagger} \chi (z) \nonumber \\
&&+  e^{-i(\overline{n}\cdot p_n n\cdot z/2
 +n\cdot p_{\bar{n}} \overline{n}\cdot z/2)}\overline{\chi}
 \overline{W} \overline{Y}^{\dagger} \gamma^{\mu} \tilde{Y} W^{\dagger}  \xi
 (z) \Bigr].
\end{eqnarray}

 Now we define the hadronic tensor $\Pi^{\mu\nu}$ as
 \begin{equation}
 \Pi^{\mu\nu} = i\int d^4 z e^{iq\cdot z} T[j^{\mu} (z) j^{\nu} (0)],
 \end{equation}
which can be written as
\begin{eqnarray}
\Pi^{\mu \nu} &=& iC^2 (Q) \int d^4 z e^{i(\bar{n}\cdot q-\bar{n}\cdot
  p_n)n\cdot z/2} e^{i(n\cdot q-n\cdot p_{\bar{n}}) \bar{n}\cdot z/2}
T\Bigl[ \overline{\chi} \overline{W}
  \overline{Y}^{\dagger} \gamma^{\mu} \tilde{Y} W^{\dagger} \xi (z)
  \overline{\xi} W \tilde{Y}^{\dagger} \gamma^{\nu}
  \tilde{\overline{Y}} \overline{W}^{\dagger} \chi (0) \Bigr]
  \nonumber \\
&=&iC^2 (Q) \int d^4 z e^{i n\cdot (q-p_{\bar{n}})
  \overline{n} \cdot z/2} T\Bigl[ \overline{\chi} \overline{W}
  \overline{Y}^{\dagger} \gamma^{\mu} \tilde{Y} W^{\dagger} \xi (z)
  \overline{\xi} W \tilde{Y}^{\dagger} \gamma^{\nu}
  \tilde{\overline{Y}} \overline{W}^{\dagger} \chi (0) \Bigr],
\end{eqnarray}
where the first exponential turns into $\delta_{Q,\bar{n}\cdot p_n}$,
$n\cdot (q-p_{\bar{n}}) = (1-x)Q$, and the perpendicular part is
omitted. Since there are no collinear particles in the final state, we
can write
\begin{equation}
\langle 0| T[W^{\dagger} \xi (z) \overline{\xi} W (0)]|0\rangle \equiv
i\frac{\FMslash{n}}{2} \int \frac{d^4 k}{(2\pi)^4} e^{-ik\cdot z} J_P
(k),
\end{equation}
where this equation defines the jet function $J_P (k)$ which depends
only on $n\cdot k$. We go down to $\mathrm{SCET}_{\mathrm{II}}$ and
relabel the usoft Wilson line $Y$,$\tilde{Y}$ as $S$, $\tilde{S}$
respectively and $\Pi^{\mu\nu}$ is written as
\begin{eqnarray}
\Pi^{\mu\nu} &=& -C^2 (Q) \int \frac{d^4 k}{(2\pi)^4} \int d^4 z
e^{i\Bigl(Q(1-x) -n\cdot k\Bigr) \overline{n} \cdot
  z/2} e^{-i   k_{\perp} \cdot z_{\perp}} e^{-i\overline{n} \cdot k
  n\cdot z/2} J_P (n\cdot k)  \\
&& \times T \Bigl[ \overline{\chi} \overline{W} \overline{S}^{\dagger}
\gamma^{\mu} \tilde{S} (z) \frac{\FMslash{n}}{2} \tilde{S}^{\dagger}
\gamma^{\nu} \tilde{\overline{S}} \overline{W}^{\dagger} \chi (0)
\Bigr] \nonumber \\
&=& -C^2 (Q) \int \frac{d\overline{n} \cdot z dn\cdot k}{4\pi}
e^{i\Bigl(Q(1-x) -n\cdot k\Bigr) \overline{n}\cdot z/2} J_P (n\cdot k)
T\Bigl[ \overline{\chi} \overline{W} \overline{S}^{\dagger}
\gamma^{\mu} \tilde{S} (z) \frac{\FMslash{n}}{2} \tilde{S}^{\dagger}
\gamma^{\nu} \tilde{\overline{S}} \overline{W}^{\dagger} \chi (0)
\Bigr] \nonumber \\
&=& C^2 (Q) g_{\perp}^{\mu\nu} T\Bigl[ \overline{\chi} \overline{W}
\frac{\FMslash{n}}{2} \overline{W}^{\dagger} \chi \Bigr] \int
\frac{d\overline{n}\cdot z dn\cdot k}{4\pi} \int d\eta
e^{i\Bigl(Q(1-x) -n\cdot k -\eta \Bigr) \overline{n}\cdot z/2} J_P
(n\cdot k) \nonumber \\
&&\times T\Bigl[\frac{1}{N} \mathrm{tr} \Bigl(\overline{S}^{\dagger}
  \tilde{S} \delta (\eta +n\cdot i\partial) \tilde{S}^{\dagger}
  \tilde{\overline{S}} \Bigr) \Bigr] \nonumber \\
&=& g_{\perp}^{\mu\nu} \int d\omega C^2 (\omega) T\Bigl[
\overline{\chi} \overline{W} 
\frac{\FMslash{n}}{2} \delta (\omega -\mathcal{P}_+)
\overline{W}^{\dagger} \chi \Bigr]  \int dn\cdot 
k J_P (n\cdot k) T\Bigl[K\Bigl( Q(1-x) -n\cdot k\Bigr) \Bigr],
\nonumber 
\end{eqnarray}
where $g_{\perp}^{\mu\nu} = g^{\mu\nu} - (n^{\mu} \overline{n}^{\nu}
+\overline{n}^{\mu} n^{\nu})/2$, and we use the fact that the soft
part is decoupled from the collinear part and only the color-singlet
component is projected out in the final result. The soft Wilson line
operator $K(\eta)$ in this case is given by
\begin{eqnarray}
K(\eta) &=& \frac{1}{N} \mathrm{tr}
\Bigl(\overline{S}^{\dagger} \tilde{S} \delta (\eta
+n\cdot i\partial) \tilde{S}^{\dagger} \tilde{\overline{S}} \Bigr)
\nonumber \\
&=&  \frac{1}{N} \mathrm{tr} \exp \Bigl[ -g\overline{n} \cdot A_s
\frac{1}{\overline{n} \cdot \mathcal{P}^{\dagger} -i\epsilon} \Bigr]
\exp \Bigl[ \frac{1}{n\cdot \mathcal{P} -i\epsilon} (-g n\cdot A_s)
\Bigr] \nonumber \\
&&\times \delta (\eta +n\cdot i\partial) \exp \Bigl[ -gn\cdot A_s
\frac{1}{n\cdot \mathcal{P}^{\dagger} +i\epsilon} \Bigr] \exp
\Bigl[\frac{1}{\overline{n} \cdot \mathcal{P} -i\epsilon}
(-g\overline{n} \cdot A_s) \Bigr].
\end{eqnarray}

\subsection{Toy $\pi$-$\gamma$ form factor}
If we consider the $\pi$-$\gamma$ form factor in the process $\gamma^*
\gamma \rightarrow \pi^0$ in which $\gamma^*$ denotes a virtual
photon, there is no soft Wilson lines of the form $K(\eta)$, as we
will explain below. However, let us consider the fictitious process
$\gamma^* (q_1) \gamma^* (q_2) \rightarrow \pi^0$, where both photons
are virtual with $q_1^{\mu} + q_2^{\mu} = Q\overline{n}^{\mu}/2$. In
this example, we only want to show that the soft
Wilson lines in this process has a different analytic structure, and
it appears in the matrix element. This may not be of physical
relevance, and it is considered as a toy calculation.

\begin{figure}[t]
\begin{center}
\epsfig{file=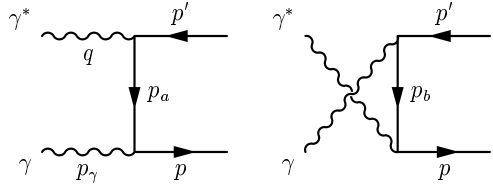, width=6.0cm}
\end{center}
\vspace{-1.cm}
\caption{Feynman diagrams contributing to the $\pi$-$\gamma$ form
  factor.\label{figpigam}}
\end{figure}

The Feynman diagrams at lowest 
order for the form factor in $\gamma^* \gamma \rightarrow \pi^0$ is
shown in Fig.~\ref{figpigam}. 
The matrix element for this transition defines the $\pi$-$\gamma$ form
factor in the full theory \cite{Bauer:2002nz}
\begin{eqnarray} \label{pgam}
  \langle \pi^0 (p_{\pi}) | j_{\mu} (0) |\gamma
(p_{\gamma},\epsilon)\rangle &=&   ie \epsilon^{\nu} \int d^4 z
e^{-ip_{\gamma} \cdot z} \langle \pi^0   (p_{\pi}) |T[j_{\mu} (0)
j_{\nu} (z)] |0\rangle \nonumber \\
&=& -ie F_{\pi\gamma} (Q^2) \epsilon_{\mu\nu\alpha\beta} p_{\pi}^{\nu}
\epsilon^{\alpha} q^{\beta},
\end{eqnarray}
where $\epsilon^{\nu}$ is the polarization vector for the real photon.
The electromagnetic current is given by $j_{\mu} =\overline{\psi}
\gamma^{\mu} \psi$. In the Breit frame the momentum of the virtual
photon $q^{\mu}$, the momentum of the real photon $p_{\gamma}^{\mu}$
and the momentum of the pion $p_{\pi}^{\mu}$ can be written as
\begin{equation}
  q^{\mu} = \frac{Q}{2} (\overline{n}^{\mu} -n^{\mu}), \
  p_{\gamma}^{\mu} = E n^{\mu} \sim \frac{Q}{2} n^{\mu}, \
  p_{\pi}^{\mu} = E_{\pi} \overline{n}^{\mu}.
\end{equation}
We consider the case where the momentum transfer from the virtual
photon is large $-q^2 =Q^2 \gg \Lambda_{\mathrm{QCD}}^2$ where
$q=p_{\pi} -p_{\gamma}$. If we write the momentum of the quark inside
a pion as $p = x p_{\pi}$ and that of the antiquark as $p^{\prime} =
-(1-x) p_{\pi}$, where $x$ is the longitudinal momentum fraction of
the pion, the virtualities of the intermediate states in
Fig.~\ref{figpigam} are given by
\begin{equation}
  p_a ^2 = -xQ^2, \ p_b^2 =- Q^2 (1-x),
\end{equation}
with $E_{\pi} \sim Q/2$ at leading order in $\Lambda_{\mathrm{QCD}}$.

Away from the endpoint region, $p_a^2 \sim p_b^2 \sim Q^2 \gg
\Lambda_{\mathrm{QCD}}^2$, and these states are integrated out to
yield effective local operators in $\mathrm{SCET}_{\mathrm{I}}$. This
is not true near the endpoint region. Near $x\sim 1$, $p_a^2$ is of
order $Q^2$, but $p_b^2 \sim (1-x) Q^2$ is small, and the intermediate
state in the second Feynman diagram of Fig.~\ref{figpigam} cannot be
integrated out in $\mathrm{SCET}_{\mathrm{I}}$. But we can
still apply SCET and the two-step matching process is useful. First we
integrate out the modes of order $p^2 \sim Q^2$ to go down to
$\mathrm{SCET}_{\mathrm{I}}$. The intermediate collinear state in the
first Feynman diagram of Fig.~\ref{figpigam} is integrated out to produce
local operators in $\mathrm{SCET}_{\mathrm{I}}$, in which the effect
of the soft Wilson line does not appear since they cancel. Now we
integrate out the modes of order $p^2 \sim (1-x)Q^2$ in the second
diagram of Fig.~\ref{figpigam} to go down to
$\mathrm{SCET}_{\mathrm{II}}$. The problem here is  that the momentum
$p^{\prime}_{\mu}$ of the antiquark becomes usoft near the endpoint
$x\sim 1$, and the effective current operator is an usoft-collinear
current. The effect of the usoft interactions near the endpoint can be
interesting on its own, but it is not appropriate in our context.  

\begin{figure}[t]
\begin{center}
\epsfig{file=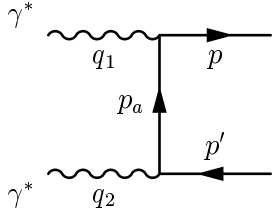, width=3.0cm}
\end{center}
\vspace{-1.cm}
\caption{Feynman diagrams for the process $\gamma^* \gamma^*
  \rightarrow \pi^0$.\label{fpigam}}
\end{figure}

Instead, let us consider a hypothetical process $\gamma^* \gamma^*
\rightarrow \pi^0$, which is shown in Fig.~\ref{fpigam}. The momenta
of the virtual photons are given by
\begin{equation}
q_1^{\mu} = \frac{Q}{2} (y\overline{n}^{\mu} -n^{\mu}), \ q_2^{\mu} =
\frac{Q}{2} \Bigl( (1-y) \overline{n}^{\mu} +n^{\mu} \Bigr),  \ \
0\leq y \leq 1, 
\end{equation}
such that $q_1^{\mu} + q_2^{\mu} = Q\overline{n}^{\mu}/2= E_{\pi}
\overline{n}^{\mu} = p_{\pi}^{\mu}$. And we write the momentum of the
quarks $p$ and $p^{\prime}$ as $p =xp_{\pi}$, $p^{\prime} = -(1-x)
p_{\pi}$. The momentum of the intermediate state is given by
\begin{equation}
p_a^{\mu} = \frac{Q}{2} \Bigl( n^{\mu} + (x-y) \overline{n}^{\mu}
\Bigr).  
\end{equation}
If $x-y \ll 1$ such that $p_a^2 = Q^2 (x-y) \sim Q\Lambda$, the
intermediate state becomes a collinear particle in the $n^{\mu}$
direction, and we can apply the two-step matching in this case. 

\begin{figure}[b]
\begin{center}
\epsfig{file=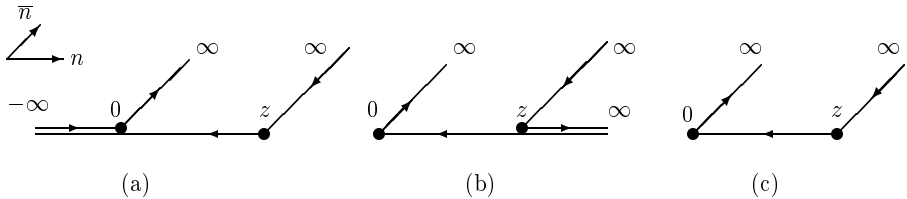, width=12.0cm}
\end{center}
\vspace{-1.cm}
\caption{The description of the soft Wilson lines in the $\pi$-$\gamma$
form factor. (a) $\xi$ from $-\infty$, (b) $\xi$ to $\infty$, (c) the
resultant soft Wilson lines from (a) and (b). \label{figpi}}
\end{figure}

Let us define the hadronic tensor
\begin{equation}
W_{\mu\nu} = i\int d^4 z e^{-i q_1 \cdot z} T\Bigl[ j_{\mu} (z)
j_{\nu} (0)\Bigr],   
\end{equation}
where the current $j_{\mu}$ is given by 
\begin{equation}
  j_{\mu} (z) = C(Q) \Bigl(e^{i(\overline{n}\cdot p_n n\cdot z/2 -n\cdot
  p_{\bar{n}} \overline{n} \cdot z/2)} \overline{\xi} W  \gamma_{\mu}
  \overline{W}^{\dagger} \chi
+e^{-i(\overline{n}\cdot p_n n\cdot z/2 -n\cdot
  p_{\bar{n}} \overline{n} \cdot z/2)}\overline{\chi} \overline{W}
  \gamma_{\mu} W^{\dagger}\xi \Bigr),
\end{equation}
where $p_n$, $p_{\bar{n}}$ are the label momenta of the corresponding
fields. Now we factorize the usoft interactions by redefining the
collinear fields. Since the collinear field $\xi$ is the intermediate
state, there are two possibilities to assign the usoft Wilson lines as
shown in Fig.~\ref{figpi} (a), and (b), which are equivalent to the
configuration in Fig.~\ref{figpi} (c).  
We follow the prescription given by Fig.~\ref{figpi} (a) and 
the collinear fields are redefined as
 \begin{eqnarray}
\overline{\xi} W \gamma^{\mu} \overline{ W}^{\dagger}  \chi : &&
\overline{\xi} \rightarrow \overline{\xi} Y^{\dagger}, \
A_n^{\mu}   \rightarrow Y A_n^{\mu}  Y^{\dagger}, \
\chi \rightarrow  \tilde{\overline{Y}}  \chi,  \ A_{\bar{n}}^{\mu}
\rightarrow  \tilde{\overline{Y}} A_{\bar{n}}^{\mu}
\tilde{\overline{Y}}^{\dagger}, \nonumber \\
\overline{\chi} \overline{W} \gamma^{\mu} W^{\dagger}
 \xi : && \overline{\chi} \rightarrow \overline{\chi}
 \tilde{\overline{Y}}^{\dagger}, \ A_{\bar{n}}^{\mu} \rightarrow
\tilde{\overline{Y}} A_{\bar{n}}^{\mu} \tilde{\overline{Y}}^{\dagger},
\ \xi   \rightarrow Y \xi,  \  A_n^{\mu} \rightarrow Y A_n^{\mu}
  Y^{\dagger}.
 \end{eqnarray}

\begin{table}[b]
  \caption{Summary of the soft Wilson lines $K(\eta)$ in various
    processes. \label{table2}}
  \centering
  \begin{tabular}{cc|cc} \hline
process&  $K(\eta)$ &process&$K(\eta)$ \\ \hline
DIS& $\displaystyle \frac{1}{N} \mathrm{tr} \Bigl(
    \tilde{\overline{S}}^{\dagger} S \delta  (\eta+n\cdot  i\partial)
    S^{\dagger}     \overline{S}   \Bigr)$ &
Drell-Yan& $\displaystyle \frac{1}{N} \mathrm{tr}
  \Bigl( \tilde{\overline{S}}^{\dagger} \tilde{S} \delta
  (\eta +n\cdot i\partial) \tilde{S}^{\dagger} \overline{S} \Bigr)$  \\
$e^+ e^-$& $\displaystyle \frac{1}{N} \mathrm{tr}
\Bigl(\overline{S}^{\dagger} \tilde{S} \delta (\eta
+n\cdot i\partial) \tilde{S}^{\dagger} \tilde{\overline{S}} \Bigr)$ &
$\pi$-$\gamma$& $\displaystyle \frac{1}{N} \mathrm{tr} \Bigl(
\tilde{\overline{S}}^{\dagger} S  \delta (\eta +n\cdot i\partial)
S^{\dagger} \tilde{\overline{S}} \Bigr)$ \\
\hline
  \end{tabular}
\end{table}

The hadronic tensor can be written as
\begin{eqnarray}
W_{\mu\nu} &=& i\int d^4 z e^{-i q_1 \cdot z} T [j_{\mu} (z)
j_{\nu} (0)] \nonumber \\
&=& i C^2 (Q)\int d^4 z  e^{i(x-y) Q \bar{n} \cdot z/2}
T\Bigl[ \overline{\chi} \overline{W}
\tilde{\overline{Y}}^{\dagger} \gamma_{\mu} Y W^{\dagger} \xi (z)
\overline{\xi} Y^{\dagger} \gamma_{\nu} \tilde{\overline{Y}}
\overline{W}^{\dagger} \chi (0) \Bigr].
\end{eqnarray}
Since there are no collinear particles in the $n^{\mu}$ direction in
the final state, we can write
\begin{equation}
\langle 0| T[W^{\dagger} \xi (z) \overline{\xi} W(0)]|0\rangle = i
\frac{\FMslash{n}}{2} \int \frac{d^4 k}{(2\pi)^4} e^{-ik\cdot z} J_P
(n\cdot k),
\end{equation}
where $J_P (n\cdot k)$ is a function of $n\cdot k$ only. We can write
$W_{\mu\nu}$ in $\mathrm{SCET}_{\mathrm{II}}$ as
\begin{eqnarray}
W_{\mu\nu} &=& -\frac{C^2 (Q)}{4\pi} \int d\overline{n} \cdot z dn\cdot k
d\eta e^{i(Q(x-y)- n\cdot k -\eta)\overline{n} \cdot z/2} J_P (n\cdot
k) \nonumber \\
&&\times T\Bigl[\overline{\chi} \overline{W}
\tilde{\overline{S}}^{\dagger} \gamma_{\mu} S \frac{\FMslash{n}}{2}
  \delta (\eta +n\cdot i\partial) S^{\dagger} \gamma_{\nu}
  \tilde{\overline{S}} 
  \overline{W}^{\dagger} \chi \Bigr] \nonumber \\
&\rightarrow& \frac{i}{4} \epsilon_{\mu\nu\alpha  \beta} n^{\alpha}
\overline{n}^{\beta} \int d\omega C^2 (\omega) T[ \overline{\chi}
\overline{W} \FMslash{n} \delta (\omega -\mathcal{P}_+) \gamma_5
\overline{W}^{\dagger} \chi] \nonumber \\
&&\times \int dn\cdot k J_P (n\cdot k) 
T\Bigl[ K\Bigl(Q(x-y) -n\cdot k \Bigr) \Bigr],
\end{eqnarray}
where we extract the collinear operator proportional to $\gamma_5$. 
Here the effect of the soft gluon emission is decoupled from the
collinear sector, and it is described by the soft Wilson line as
\begin{eqnarray}
K(\eta)&=&  \frac{1}{N} \mathrm{tr}\  \tilde{\overline{S}}^{\dagger} S
\delta (\eta +n\cdot i\partial) S^{\dagger} \tilde{\overline{S}}
\nonumber \\
&=& \exp \Bigl[-gn\cdot A_s \frac{1}{\overline{n}\cdot
  \mathcal{P}^{\dagger} +i\epsilon} \Bigr] \cdot \exp  \Bigl[
\frac{1}{n\cdot \mathcal{P} +i\epsilon} (-g n\cdot A_s) \Bigr]  \delta
(\eta +n\cdot i\partial) \nonumber \\
&&\times \exp \Bigl[ -gn\cdot A_s \frac{1}{n\cdot
  \mathcal{P}^{\dagger} -i\epsilon} \Bigr] \cdot \exp \Bigl[
\frac{1}{\overline{n} \cdot \mathcal{P} -i\epsilon} (-g\overline{n}
\cdot A_s) \Bigr].
\end{eqnarray}
The forms of the soft Wilson lines for all the processes are
summarized in Table~\ref{table2}.

\section{Radiative corrections for the soft Wilson loop\label{sec4}}
As we have seen in the previous section, the effects of the soft gluon
emission expressed by the soft Wilson lines can be important near the
boundary of the phase space. The matrix element of the soft Wilson
line $K(\eta)$ between the vacuum state describes a nonperturbative
effect. Away from the boundary of the phase space, the soft Wilson
lines cancel \cite{Bauer:2002nz} or the radiative corrections can vanish
\cite{Bauer:2003di,Manohar:2003vb}. In
Refs.~\cite{Bauer:2003di,Manohar:2003vb}, the authors considered the
region where $\eta \gg n\cdot i\partial \sim \Lambda$ and the delta
function  $\delta (\eta + n\cdot i\partial)$ can be expanded using the
multipole expansion. However, we consider the region $\eta \sim n\cdot
i\partial \sim \Lambda$, where the multipole expansion is not
useful. And if we include quantum corrections, the matrix elements and
the radiative corrections can be nonzero.

\begin{table}[b]
  \begin{tabular}{cl}
 & \epsfig{file=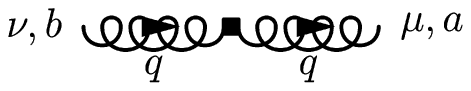, width=4.0cm} \\ \hline
DIS:&$\displaystyle g^2 \frac{\delta_{ab}}{2N}
 \Bigl[ \Bigl(
\frac{\overline{n}^{\mu} n^{\nu}}{(\overline{n}\cdot
  q+i\epsilon)(n\cdot q +i\epsilon)} +\frac{n^{\mu}
  \overline{n}^{\nu}}{(n\cdot q -i\epsilon)(\overline{n}\cdot q
  +i\epsilon)} \Bigr) \delta (\eta) $\\
&$\displaystyle +\Bigl(\frac{\overline{n}^{\mu}
  n^{\nu}}{(\overline{n}\cdot q
  +i\epsilon)(-n\cdot q-i\epsilon)} +\frac{n^{\mu}
  \overline{n}^{\nu}}{(-n \cdot q + i\epsilon)(\overline{n}\cdot q
  +i\epsilon)} \Bigr) \delta (\eta +n\cdot q)$ \\
&  $\displaystyle + \Bigl[ a\leftrightarrow b, \mu
\leftrightarrow \nu,
  q\leftrightarrow -q \Bigr] $\\ \hline
Drell-Yan:& $\displaystyle g^2 \frac{\delta_{ab}}{2N}
 \Bigl[ \Bigl(
\frac{\overline{n}^{\mu} n^{\nu}}{(\overline{n}\cdot
  q+i\epsilon)(n\cdot q -i\epsilon)} +\frac{n^{\mu}
  \overline{n}^{\nu}}{(n\cdot q +i\epsilon)(\overline{n}\cdot q
  +i\epsilon)} \Bigr) \delta (\eta) $\\
&$\displaystyle +\Bigl(\frac{\overline{n}^{\mu}
  n^{\nu}}{(\overline{n}\cdot q
  +i\epsilon)(-n\cdot q+i\epsilon)} +\frac{n^{\mu}
  \overline{n}^{\nu}}{(-n \cdot q - i\epsilon)(\overline{n}\cdot q
  +i\epsilon)} \Bigr) \delta (\eta +n\cdot q)$ \\
&  $\displaystyle + \Bigl[ a\leftrightarrow b, \mu
\leftrightarrow \nu,
  q\leftrightarrow -q \Bigr] $\\ \hline
$e^+ e^-$: & $\displaystyle g^2 \frac{\delta_{ab}}{2N}
 \Bigl[ \Bigl(
\frac{\overline{n}^{\mu} n^{\nu}}{(\overline{n}\cdot
  q-i\epsilon)(n\cdot q -i\epsilon)} +\frac{n^{\mu}
  \overline{n}^{\nu}}{(n\cdot q +i\epsilon)(\overline{n}\cdot q
  -i\epsilon)} \Bigr) \delta (\eta) $\\
&$\displaystyle +\Bigl(\frac{\overline{n}^{\mu}
  n^{\nu}}{(\overline{n}\cdot q
  -i\epsilon)(-n\cdot q +i\epsilon)} +\frac{n^{\mu}
  \overline{n}^{\nu}}{(-n \cdot q - i\epsilon)(\overline{n}\cdot q
  -i\epsilon)} \Bigr) \delta (\eta +n\cdot q)$ \\
&  $\displaystyle + \Bigl[ a\leftrightarrow b, \mu
\leftrightarrow \nu,
  q\leftrightarrow -q \Bigr] $\\ \hline
$\pi$-$\gamma$: & $\displaystyle g^2 \frac{\delta_{ab}}{2N} \Bigl[ \Bigl(
\frac{\overline{n}^{\mu} n^{\nu}}{(\overline{n}\cdot
  q+i\epsilon)(n\cdot q +i\epsilon)} +\frac{n^{\mu}
  \overline{n}^{\nu}}{(\overline{n}\cdot q -i\epsilon)(n\cdot q
  -i\epsilon)} \Bigr) \delta (\eta) $\\
&$\displaystyle +\Bigl(\frac{\overline{n}^{\mu}
  n^{\nu}}{(\overline{n}\cdot q
  +i\epsilon)(-n\cdot q-i\epsilon)} +\frac{n^{\mu}
  \overline{n}^{\nu}}{(\overline{n} \cdot q - i\epsilon)(-n\cdot q
  +i\epsilon)} \Bigr) \delta (\eta +n\cdot q)$ \\
&  $\displaystyle + \Bigl[ a\leftrightarrow b, \mu
\leftrightarrow \nu,
  q\leftrightarrow -q \Bigr] $\\  \hline
  \end{tabular}
  \caption{Feynman rules for the soft Wilson operator with two
    external gluons for deep inelastic scattering (DIS), the Drell-Yan
    process, the jet production from $e^+ e^-$ collisions, and the
    $\pi$-$\gamma$ form factor.  Only
    the terms with one $n$ and one $\overline{n}$ are shown. The arrows
indicate the momentum flow.\label{sfeyn}}
\label{table3}
 \end{table}

\begin{figure}[t]
\begin{center}
\epsfig{file=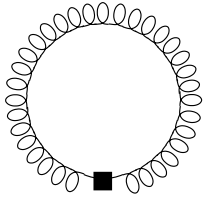,width=3.cm}
\end{center}
\vspace{-0.5cm}
\caption{Radiative correction for the soft Wilson line $K(\eta)$ at
  one loop.}
\label{figsloop}
\end{figure}

In order to compute the anomalous dimension of the soft Wilson line,
we consider the radiative corrections induced from the operator
$K(\eta)$, and the Feynman rules for the two-point vertex in various
processes is given in Table~\ref{table3}. The Feynman diagram for the
radiative correction at one loop is shown in Fig.~\ref{figsloop} and, in
deep inelastic scattering, it is given by
\begin{eqnarray} \label{disr}
  I_{\mathrm{DIS}} (\eta) &=&-ig^2 C_F \int \frac{d^D l}{(2\pi)^D}
\Bigl[\frac{4\delta(\eta)}{ l^2 (\overline{n} \cdot l
  -\lambda_1)(n\cdot l -\lambda_2)} \nonumber \\
&+& \frac{\delta (\eta  +n\cdot l)}{(l^2 +i0)
  (\overline{n}\cdot l -\lambda_1 +i0)} \Bigl( \frac{1}{-n\cdot l
    +\lambda_2   -i0} + \frac{1}{-n\cdot l  +\lambda_2   +i0}
  \Bigr)\nonumber \\ 
&+&\frac{\delta (\eta -n\cdot l)}{(l^2 +i0)(-\overline{n} \cdot l
  -\lambda_1 +i0)} \Bigl(\frac{1}{n\cdot l +\lambda_2 -i0}
  +\frac{1}{n\cdot l +\lambda_2 +i0} \Bigr) \Bigr],
\end{eqnarray}
where $\lambda_1$, and $\lambda_2$ are the infrared cutoffs, and
$i\epsilon$ is replaced by $i0$ to avoid confusion. Here the $i0$
prescription is omitted in the first integral including $\delta
(\eta)$ because it does not affect the calculation. In order to
extract the ultraviolet divergence as 
poles in $1/\epsilon$, we employ dimensional regularization with
$D=4-2\epsilon$, and we introduce infrared cutoffs to regulate
infrared divergences. A rigorous method to introduce the infrared
cutoff without violating the gauge invariance is discussed in
Ref.~\cite{Bauer:2003td}, and it should be followed. But here we put
simple infrared cutoffs in order to extract the ultraviolet divergence
only and to see that there is no mixing of the ultraviolet and
infrared divergences.  

The integral proportional to $\delta (\eta)$ in Eq.~(\ref{disr}) is
given by 
\begin{eqnarray}\label{idisa}
  I_{\mathrm{DIS}}^a (\eta) &=& -4ig^2C_F \delta (\eta) \int \frac{d^D
  l}{(2\pi)^D}   \frac{1}{l^2 (\overline{n} \cdot l-\lambda_1)( n\cdot
  l -\lambda_2)}   \nonumber \\
&=&  -4ig^2C_F \delta (\eta) 8\int_0^{\infty} du \int_0^{\infty} dv \int
  \frac{d^D l}{(2\pi)^4} \frac{1}{\Bigl[l^2 -(4uv +2u\lambda_1
  +2v\lambda_2) \Bigr]^3} \nonumber \\
&=& -\frac{\alpha_s C_F}{\pi} \Bigl(\frac{\lambda_1 \lambda_2}{\mu^2}
  \Bigr)^{-\epsilon} \delta (\eta)
 \frac{1}{\epsilon^2},
\end{eqnarray}
where only the ultraviolet divergent term is kept.

For other integrals, we first integrate over $n\cdot l$ with the delta
functions, and the remaining integral is computed using the contour
integral in the complex $\overline{n} \cdot l$ plane, and we finally
integrate over $l_{\perp}$ using the dimensional regularization in
$D-2(=2-2\epsilon)$ dimensions. Since the integration of the delta
functions $\delta (\eta \pm n\cdot l)$ over $n\cdot l$ produces a
number, the $i0$ prescription does not affect the computation, while
the position of the poles in the complex $\overline{n} \cdot l$ plane
becomes important. Therefore the two integrands proportional to
$\delta (\eta +n\cdot l)$ and those proportional to  $\delta (\eta
-n\cdot l)$, which differ only in 
the $i0$ prescription in the terms containing $n\cdot l$, give the
same result. That is, the analytic structure of the soft Wilson lines
in the $n^{\mu}$ directions does not alter the radiative corrections
when the separation between 0 and $z$ is lightlike in the $n^{\mu}$
direction. As long as the analytic structure is the same for the soft
Wilson lines in the $\overline{n}^{\mu}$ direction, the radiative
corrections for the soft Wilson lines are the same though different
specifications of the soft Wilson lines in the $n^{\mu}$ direction may 
correspond to different physical processes. For this reason, the 
configuration of the soft Wilson lines in the Drell-Yan process shown
in Fig.~\ref{figdy} (a) gives the same radiative corrections as the
configuration give in Fig.~\ref{figdy} (b) as long as the separation
is lightlike.

The integral with $\delta (\eta +n\cdot l)$ is given by
\begin{eqnarray}
  I_{\mathrm{DIS}}^b  (\eta)&=& -2ig^2 C_F \int \frac{d^D l}{(2\pi)^D}
  \frac{1}{\overline{n}\cdot  l n\cdot l -l_{\perp}^2 +i0}
  \frac{1}{\overline{n} \cdot l
  -\lambda_1 +i0} \frac{1}{-n\cdot l+\lambda_2 -i0} \delta
  (\eta +n\cdot l) \nonumber \\
&=& ig^2 C_F \frac{1}{\eta +\lambda_2} \int \frac{d^{D-2}l_{\perp}
  d\overline{n} \cdot l}{(2\pi)^D} \frac{1}{-\overline{n} \cdot l \eta
  -l_{\perp}^2 +i0} \frac{1}{\overline{n} \cdot
  l-\lambda_1+i0}.
\end{eqnarray}
In order for this integral to have nonzero value, the poles in the
complex $\overline{n}\cdot l$ plane should be on the opposite side of
the real axis, hence $\eta$ should be positive, and the
integral is given by
\begin{equation}
  I_{\mathrm{DIS}}^b (\eta)
  =\frac{\alpha_s C_F}{2\pi}   \Gamma   (\epsilon) \theta (\eta)
  \frac{(\eta   \lambda_1   /\mu^2)^{-\epsilon}}{\eta   +\lambda_2}.
\end{equation}
Similarly, the integral with $\delta (\eta -n\cdot l)$ is given by
\begin{equation}
  I_{\mathrm{DIS}}^c (\eta) = \frac{\alpha_s C_F}{2\pi}   \Gamma
  (\epsilon)   \theta (\eta)   \frac{(\eta
  \lambda_1/\mu^2)^{-\epsilon}}{\eta   +\lambda_2},
\end{equation}
which is the same as $I_{\mathrm{DIS}}^b (\eta)$. In order to extract
the ultraviolet divergence, we can  write $I_{\mathrm{DIS}}^b (\eta)$
in the form
\begin{equation}
 I_{\mathrm{DIS}}^b (\eta) = A\delta (\eta) + B\theta (\eta) \Bigl(
 \frac{1}{\eta} \Bigr)_+,
\end{equation}
since the singularity resides at $\eta=0$ and the above equation
defines the ``+''-distribution which satisfies
\begin{equation}
\int_{-\infty}^{\infty} d\eta \, \theta (\eta) \Bigl(\frac{1}{\eta}
  \Bigr)_+ =0, \ \int_{-\infty}^{\infty} d\eta \, \theta (\eta)
  f(\eta)   \Bigl(\frac{1}{\eta}   \Bigr)_+ = \int_0^{\infty} d\eta
  \frac{f(\eta) -f(0)}{\eta},
\end{equation}
where $f(\eta)$ is a regular function at $\eta=0$. Determining $A$
and $B$ by explicit calculation, we obtain
\begin{equation}
  I_{\mathrm{DIS}}^{b,c} (\eta) 
= \frac{\alpha_s C_F}{2\pi} \Bigl[\Bigl(\frac{\lambda_1
  \lambda_2}{\mu^2}\Bigr)^{-\epsilon} \frac{1}{\epsilon^2} \delta
  (\eta) +\frac{1}{\epsilon} \frac{\theta(\eta)}{(\eta)_+}\Bigr].
\end{equation}
By adding all the contributions, the result is given by
\begin{equation}
  I_{\mathrm{DIS}} (\eta) =   I_{\mathrm{DIS}}^a  (\eta) +
  I_{\mathrm{DIS}}^b (\eta) +I_{\mathrm{DIS}}^c (\eta) =
  \frac{\alpha_s C_F}{\pi} \frac{1}{\epsilon} \frac{\theta
  (\eta)}{(\eta)_+}.
\end{equation}

The relation between the bare operator and the renormalized operator
can be written, in general, as
\begin{equation}
  K_B (\eta) = \int d\eta^{\prime} Z_{\mathrm{DIS}} (\eta,
  \eta^{\prime}) K_R
  (\eta^{\prime}),
\end{equation}
where $Z(\eta,\eta^{\prime})$ is given by
\begin{equation}
  Z_{\mathrm{DIS}} (\eta, \eta^{\prime}) = \delta (\eta-\eta^{\prime})
  +\frac{\alpha_s C_F}{\pi}
  \frac{1}{\epsilon}
  \frac{\theta (\eta -\eta^{\prime})}{(\eta -\eta^{\prime})_+}.
\end{equation}
Therefore the renormalization group equation for the renormalized
operator is given by
\begin{equation} \label{rgdis}
  \Bigl(\mu \frac{\partial}{\partial \mu} +\beta
  \frac{\partial}{\partial g} \Bigr) K (\eta) = -\int d\eta^{\prime}
  \gamma_{\mathrm{DIS}} (\eta, \eta^{\prime})  K(\eta^{\prime}),
\end{equation}
where the anomalous dimension $\gamma_{\mathrm{DIS}}
(\eta,\eta^{\prime})$ is given by
\begin{equation} \label{gdis}
  \gamma_{\mathrm{DIS}} (\eta ,\eta^{\prime}) = -2\frac{\alpha_s
  C_F}{\pi}  \frac{\theta (\eta-\eta^{\prime})}{(\eta
  -\eta^{\prime})_+}.
\end{equation}

Now let us consider the radiative correction for the soft gluon
emission in the Drell-Yan process. The Feynman diagram is also given by
Fig.~\ref{figsloop}, but with the different Feynman rule in
Table~\ref{sfeyn} for the Drell-Yan process. The radiative
correction at one loop is given by
\begin{eqnarray} \label{drr}
  I_{\mathrm{DY}} (\eta) &=&-2ig^2 C_F \int \frac{d^D l}{(2\pi)^D}
\Bigl[\frac{2\delta(\eta)}{ l^2 (\overline{n} \cdot l
  -\lambda_1)(n\cdot l -\lambda_2)} \nonumber \\
&+& \frac{\delta (\eta  +n\cdot l)}{(l^2 +i0)
  (\overline{n}\cdot l -\lambda_1 +i0)(-n\cdot l +\lambda_2
  +i0)} \nonumber \\
&+&\frac{\delta (\eta -n\cdot l)}{(l^2 +i0)(-\overline{n} \cdot l
  -\lambda_1 +i0)(n\cdot l +\lambda_2 +i0)} \Bigr].
\end{eqnarray}
Compared to the case of deep inelastic scattering in Eq.~(\ref{disr}),
the only difference is the $i0$ prescription in the $n\cdot l$ part in
the denominators of the second and the third lines in
Eq.~(\ref{drr}). Since the integration over $n\cdot l$ is governed by
the delta functions $\delta (\eta \pm n\cdot l)$, the $i0$
prescription does not have any effect on the result of the integration
and the radiative correction for the Drell-Yan process is the same
as the case in deep inelastic scattering. And the renormalization
group equation for the soft Wilson line in Drell-Yan process is given
by
\begin{equation}
  \Bigl(\mu \frac{\partial}{\partial \mu} +\beta
  \frac{\partial}{\partial g} \Bigr) K (\eta) = -\int d\eta^{\prime}
  \gamma_{\mathrm{DY}} (\eta, \eta^{\prime})  K(\eta^{\prime}),
\end{equation}
which is the same as that of deep inelastic scattering in
Eq.~(\ref{rgdis})  with the same anomalous dimension
$\gamma_{\mathrm{DY}} (\eta, \eta^{\prime} ) =\gamma_{\mathrm{DIS}}
(\eta, \eta^{\prime})$. 

This result is consistent with the previous results on the resummation
of large infrared corrections of the Wilson loops on the light cone in
Refs.~\cite{Korchemsky:1993uz}. In order to see
the consistency, let us begin with the renormalization group equation,
Eq.~(4.5) for the Wilson loop on the light cone in coordinate space in
Ref.~\cite{Korchemsky:1993uz}, which is given by
\begin{equation}
\Bigl( \mu\frac{\partial}{\partial \mu} +\beta (g)
\frac{\partial}{\partial g} \Bigr) W (\rho -i0) = -\Bigl[
\Gamma_{\mathrm{cusp}} (g) \Bigl( \ln (\rho -i0) +\ln (-\rho
+i0)\Bigr) +\Gamma (g) \Bigr] W(\rho -i0),
\end{equation}
where $\Gamma (g)$ is the integration constant, $\rho =(v\cdot y)
\mu$, and  $\Gamma_{\mathrm{cusp}}$ is called the cusp anomalous
dimension \cite{korchem87}.
This is the renormalization group equation for the soft Wilson
loop with the initial quark with velocity $v^{\mu}$. However, we can
put the initial particle on the light cone, say, in the
$\overline{n}^{\mu}$ direction. In this case, the integration constant
becomes $\Gamma (g) = 0+O(\alpha_s^2)$. At order $\alpha_s$, the
renormalization group equation for the soft Wilson loop becomes
\begin{equation} \label{swilson}
\Bigl( \mu\frac{\partial}{\partial \mu} +\beta (g)
\frac{\partial}{\partial g} \Bigr) W (\rho -i0) = -\Bigl[
\Gamma_{\mathrm{cusp}} (g) \Bigl( \ln (\rho -i0) +\ln (-\rho
+i0)\Bigr) +\Gamma (g) \Bigr] W(\rho -i0),
\end{equation}
where $\rho = (\overline{n} \cdot y) \mu$, and the $i0$ prescription
comes from the Feynman prescription.

Instead of considering $W$ as a function of $\rho$, we will consider
$W$ as a function of $\overline{n}\cdot y$, $W(\rho) =
W(\overline{n}\cdot y, \mu, g)$ since its Fourier transform is what we
have computed so far.
Let us define the Fourier transform of $W(\overline{n} \cdot y)$ as
\begin{equation}
W(\eta)= \int \frac{d\overline{n} \cdot y}{2\pi} e^{i\overline{n}\cdot
   y\eta} W(\overline{n}\cdot y -i0), \
 W(\overline{n}\cdot y -i0) =  \int d\eta e^{-i\overline{n}\cdot y
   \eta} W (\eta),
\end{equation}
and we take the Fourier transform of Eq.~(\ref{swilson}). It becomes
\begin{eqnarray} \label{kore}
&&\Bigl( \mu\frac{\partial}{\partial \mu} +\beta (g)
\frac{\partial}{\partial g} \Bigr) W (\eta) \nonumber \\
&& =
 - \Gamma_{\mathrm{cusp}} \int \frac{d\overline{n} \cdot y}{2\pi}
e^{i\overline{n} \cdot y \eta}
\Bigl( \ln (\overline{n} \cdot y \mu -i0) +\ln (-\overline{n}\cdot y \mu
+i0)\Bigr)
W(\overline{n}\cdot y \mu -i0)  \nonumber \\
&&= - \Gamma_{\mathrm{cusp}} \int \frac{d\overline{n} \cdot y}{2\pi}
e^{i\overline{n} \cdot y \eta} \Bigl( \ln (\overline{n} \cdot y \mu
-i0) +\ln (-\overline{n}\cdot y \mu +i0)\Bigr) \int d\eta^{\prime}
e^{-i\overline{n} \cdot y \eta^{\prime}} W(\eta^{\prime}) \nonumber \\
&&=  - \Gamma_{\mathrm{cusp}} \int d\eta^{\prime}  W(\eta^{\prime})
\int \frac{d\overline{n} \cdot y}{2\pi} e^{i \overline{n} \cdot y
  (\eta -\eta^{\prime})}  \Bigl( \ln (\overline{n} \cdot y \mu
-i0) +\ln (-\overline{n}\cdot y \mu +i0)\Bigr) \nonumber \\
&&=  - \Gamma_{\mathrm{cusp}} \int d\eta^{\prime} V(\eta
-\eta^{\prime})W(\eta^{\prime}),
\end{eqnarray}
where $V (\eta -\eta^{\prime})$ is given by
\begin{equation}
V (\eta -\eta^{\prime}) =  \int \frac{d\overline{n} \cdot y}{2\pi}
  e^{i \overline{n} \cdot y
  (\eta -\eta^{\prime})}  \Bigl( \ln (\overline{n} \cdot y \mu
-i0) +\ln (-\overline{n}\cdot y \mu +i0)\Bigr) 
= -2\frac{\theta (\eta -\eta^{\prime})}{(\eta -\eta^{\prime})_+}.
\end{equation}
Since $\Gamma_{\mathrm{cusp}} = \alpha_s C_F/\pi + O(\alpha_s^2)$, the
anomalous dimensions in Eq.~(\ref{kore}) and in Eq.~(\ref{gdis}) are
the same.

In the jet production from $e^+ e^-$ collisions, the radiative
correction is given by
\begin{eqnarray} \label{jetr}
  I_{\mathrm{jet}} (\eta) &=&-2ig^2 C_F \int \frac{d^D l}{(2\pi)^D}
\Bigl[\frac{2\delta(\eta)}{ l^2 (\overline{n} \cdot l
  -\lambda_1)(n\cdot l -\lambda_2)} \nonumber \\
&+& \frac{\delta (\eta  +n\cdot l)}{(l^2 +i0)
  (\overline{n}\cdot l -\lambda_1 -i0)(-n\cdot l +\lambda_2
  +i0)} \nonumber \\
&+&\frac{\delta (\eta -n\cdot l)}{(l^2 +i0)(-\overline{n} \cdot l
  -\lambda_1 -i0)(n\cdot l +\lambda_2 +i0)} \Bigr].
\end{eqnarray}
As in the previous calculations, the $i0$ prescription for the $n\cdot
l$ part does not matter, but the $i0$ prescription for the
$\overline{n} \cdot l$ part has opposite signs compared to the cases
in deep inelastic scattering and the Drell-Yan process. The integration
of the term proportional to $\delta (\eta)$ is the same, and the
integrals $I_{\mathrm{jet}}^b$ and $I_{\mathrm{jet}}^c$  with the
delta functions $\delta (\eta \pm n\cdot l)$ are given by
\begin{eqnarray}
I_{\mathrm{jet}}^b &=&  -2ig^2 C_F \int \frac{d^D l}{(2\pi)^D}
  \frac{1}{\overline{n}\cdot  l n\cdot l -l_{\perp}^2 +i0}
  \frac{1}{\overline{n} \cdot l
  -\lambda_1 -i0} \frac{1}{-n\cdot l+\lambda_2 +i0} \delta
  (\eta +n\cdot l) \nonumber \\
&=& \frac{\alpha_s C_F}{2\pi} \Bigl[\Bigl(\frac{\lambda_1
  \lambda_2}{\mu^2}\Bigr)^{-\epsilon} \frac{1}{\epsilon^2} \delta
  (\eta) +\frac{1}{\epsilon} \frac{\theta(-\eta)}{(-\eta)_+}\Bigr] =
  I_{\mathrm{jet}}^c.
\end{eqnarray}
Combining with the part proportional to $\delta (\eta)$, the
$1/\epsilon^2$ pole cancels and the radiative correction at one loop
in $e^+e^-$ collisions is given by
\begin{equation}
I_{\mathrm{jet}} = \frac{\alpha_s C_F}{\pi}
\frac{\theta(-\eta)}{(-\eta)_+}.
\end{equation}
The relation between the bare operator $K_B$ and the renormalized
operator $K_R$ is written as
\begin{equation}
K_B (\eta) = \int d\eta^{\prime} Z_{\mathrm{jet}} (\eta,
\eta^{\prime}) K_R (\eta^{\prime}),
\end{equation}
where $Z_{\mathrm{jet}} (\eta, \eta^{\prime})$ is given by
\begin{equation}
 Z_{\mathrm{jet}} (\eta, \eta^{\prime}) =\delta (\eta -\eta^{\prime})
 + \frac{\alpha_s C_F}{\pi}
\frac{\theta(\eta^{\prime}-\eta)}{(\eta^{\prime}-\eta)_+}.
\end{equation}
And the renormalized operator satisfies the renormalization group
equation
\begin{equation} \label{rgjet}
   \Bigl(\mu \frac{\partial}{\partial \mu} +\beta
  \frac{\partial}{\partial g} \Bigr) K (\eta) = -\int d\eta^{\prime}
  \gamma_{\mathrm{jet}}  (\eta, \eta^{\prime})  K(\eta^{\prime}),
\end{equation}
where the anomalous dimension $\gamma_{\mathrm{jet}}
(\eta,\eta^{\prime})$ is given by
\begin{equation} \label{gjet}
  \gamma_{\mathrm{jet}} (\eta ,\eta^{\prime}) = -2\frac{\alpha_s
  C_F}{\pi}  \frac{\theta (\eta^{\prime}-\eta)}{(\eta^{\prime}
  -\eta)_+}.
\end{equation}
When we compare the kernel, or the anomalous dimension of deep
inelastic scattering and the Drell-Yan process in Eq.~(\ref{rgdis}), the
anomalous dimension has the same form, but the argument changes sign.

Let us finally consider the radiative correction for the soft Wilson
line in $\pi$-$\gamma$ form factor. The radiative correction at one
loop in given by
\begin{eqnarray} \label{pir}
I_{\pi} (\eta) &=&  -ig^2 C_F \int \frac{d^D l}{(2\pi)^D} \frac{1}{l^2
  +i0} \Bigl[ \frac{2\delta (\eta)}{(\overline{n} \cdot l -\lambda_1)(
  n\cdot l -\lambda_2)}  \\
&+& \delta (\eta +n\cdot l) \Bigl\{ \frac{1}{(\overline{n} \cdot l
  -\lambda_1 +i0)(-n\cdot l + \lambda_2 -i0)} +\frac{1}{(\overline{n}
  \cdot l -\lambda_1 -i0)(-n\cdot l +\lambda_2 +i0)} \Bigr\} \nonumber
  \\
&+& \delta (\eta-n\cdot l) \Bigl\{ \frac{1}{(-\overline{n} \cdot l
  -\lambda_1 +i0)(n\cdot l +\lambda_2 -i0)} +\frac{1}{(-\overline{n}
  \cdot l -\lambda_1 -i0)(n\cdot l + \lambda_2 +i0)}
  \Bigr\}\Bigr]. \nonumber
\end{eqnarray}
In Eq.~(\ref{pir}), the integral which contains $\delta (\eta)$ is the
same integral which appears previously and is given by
\begin{equation}
I_{\pi}^a (\eta) = ig^2 C_F \int \frac{d^D l}{(2\pi)^D} \frac{1}{l^2
  +i0} \frac{2\delta (\eta)}{(\overline{n} \cdot l -\lambda_1)(
  n\cdot l -\lambda_2)} = -\frac{\alpha_s C_F}{2\pi}
  \Bigl(\frac{\lambda_1 \lambda_2}{\mu^2}
  \Bigr)^{-\epsilon} \delta (\eta)
 \frac{1}{\epsilon^2}
\end{equation}
Using the same technique, the integrals $I_{\pi}^b (\eta) $ and
$I_{\pi}^c (\eta)$ in
the second and the third line of Eq.~(\ref{pir}) can be evaluated and
they are given as
\begin{equation}
I_{\pi}^b (\eta) =I_{\pi}^c (\eta) = \frac{\alpha_s C_F}{4\pi}
\Gamma (\epsilon) (\eta \lambda_1)^{-\epsilon} \frac{\theta (\eta)
-\theta
  (-\eta)}{\eta +\lambda_2}.
\end{equation}
By adding all the contributions, the radiative correction for the
soft Wilson line in $\pi$-$\gamma$ form factor is given by
\begin{equation}
I_{\pi} (\eta) = I_{\pi}^a (\eta) +I_{\pi}^b (\eta) +I_{\pi}^c
(\eta) = \frac{\alpha_s C_F}{2\pi} \frac{1}{\epsilon} \frac{\theta
(\eta) -\theta
  (-\eta)}{\eta}.
\end{equation}
The relation between the bare operator $K_B (\eta)$ and the
renormalized operator $K_R (\eta)$ can be written as
\begin{equation}\label{pgcou}
K_B (\eta) =\int d\eta^{\prime} Z_{\pi\gamma} (\eta,
\eta^{\prime}) K_R (\eta^{\prime}),
\end{equation}
where $Z_{\pi\gamma} (\eta, \eta^{\prime})$ is given by
\begin{equation}
Z_{\pi\gamma} (\eta, \eta^{\prime}) = \delta (\eta -\eta^{\prime}) +
\frac{\alpha_s C_F}{2\pi} \frac{1}{\epsilon} 
\Bigl[\frac{\theta (\eta -\eta^{\prime})}{(\eta -\eta^{\prime})_+}
  +\frac{\theta (\eta^{\prime} -\eta)}{(\eta^{\prime} -\eta)_+}
  \Bigr].   
\end{equation}
The renormalized soft Wilson-line operator satisfies the
renormalization group equation
\begin{equation} \label{rgpig}
   \Bigl(\mu \frac{\partial}{\partial \mu} +\beta
  \frac{\partial}{\partial g} \Bigr) K (\eta) = -\int d\eta^{\prime}
  \gamma_{\pi\gamma}  (\eta, \eta^{\prime})  K(\eta^{\prime}),
\end{equation}
where the anomalous dimension $\gamma_{\pi\gamma}
(\eta,\eta^{\prime})$ is given by
\begin{equation} \label{gpig}
  \gamma_{\pi\gamma} (\eta ,\eta^{\prime}) = -\frac{\alpha_s
  C_F}{\pi}  \Bigl[ \frac{\theta(\eta-\eta^{\prime})}{(\eta
  -\eta^{\prime})_+} +\frac{\theta (\eta^{\prime}-\eta)}{(\eta^{\prime}
  -\eta)_+} \Bigr].  
\end{equation}

This result gives the Brodsky-Lepage kernel, and is also
consistent with the result in Ref.~\cite{Korchemskaya:1992je}. The
authors considered a soft Wilson loop with four cusps, so there is
a difference of factor 2 in front of $\Gamma_{\mathrm{cusp}}$ in
the renormalization group equation. If we perform the same
analysis for the case of deep inelastic scattering, the result, in
terms of $W(\eta)$ can be written as
\begin{eqnarray}
\Bigl( \mu\frac{\partial}{\partial \mu}
+\beta\frac{\partial}{\partial
  g} \Bigr) W(\eta, g) &=& -\Gamma_{\mathrm{cusp}} \int d\eta^{\prime}
  V(\eta -\eta^{\prime}) W(\eta^{\prime},g) \nonumber \\
&=& \frac{\alpha_s C_F}{\pi} \int d\eta^{\prime}  \Bigl(
\frac{\theta
  (\eta-\eta^{\prime})}{(\eta-\eta^{\prime})_+} +
  \frac{\theta (\eta^{\prime} -\eta)}{(\eta^{\prime} -\eta)_+} \Bigr)
  W(\eta^{\prime}, g).
\end{eqnarray}
This renormalization group equation is exactly the same as
Eqs.~(\ref{rgpig}), and (\ref{gpig}).

\section{Conclusion\label{sec5}}
We have considered the effect of the  soft gluon emission in
high-energy processes near the phase boundaries. The soft gluon
emission can be expressed in terms of the soft Wilson-line
operators. Since the collinear and 
energetic particles are decoupled from the soft interactions in
$\mathrm{SCET}_{\mathrm{II}}$, the matrix element of the soft Wilson
lines is shown to be factorized. Therefore the decay rates or the
scattering cross sections can be written as a convolution of the
Wilson coefficients, the jet functions, the matrix elements of the
collinear operators, and the matrix elements of the soft Wilson
lines. The separation of long-distance physics and short-distance
physics is complicated in the full theory, and SCET offers a simple
tool to see this factorization clearly. And the scaling behavior of
each factorized part can be computed in perturbation theory order by
order. In this paper, we have focused on the scaling behavior of the
soft Wilson lines. 

The appearance of the soft Wilson line is
universal when we consider the effect of the soft gluon emission near
the phase boundary. But the analytic structure of the soft Wilson
line is different in different processes, which appear as the
$i\epsilon$ prescription in the factorized exponential, or the path-
and anti path-ordering of the gauge fields in the soft Wilson lines. At
tree level, the difference of the analytic structure in the soft
Wilson lines may not appear and the effect of the soft gluon emission
is truly universal. If we include quantum corrections, the
different analytic structure affects the scaling behavior of the soft
Wilson lines and it has been explicitly shown in this paper by
computing the anomalous dimensions of the soft Wilson lines in various
processes.  The anomalous dimensions for the soft Wilson line in deep
inelastic scattering and in the Drell-Yan process are the same, while
the anomalous dimension in the jet production from $e^+ e^-$
collisions has the same form but the sign of the argument is
opposite. In the $\pi$-$\gamma$ form factor, the anomalous dimension
is proportional to $\theta (\eta-\eta^{\prime})/(\eta-\eta^{\prime})_+
+\theta (\eta^{\prime}-\eta)/(\eta^{\prime}-\eta)_+$,  and this gives
the Brodsky-Lepage kernel. This originates from the different analytic
structure of the soft Wilson lines in the $\overline{n}^{\mu}$
direction for the lightlike separation in the $n^{\mu}$ direction.

Another advantage in adopting SCET in considering the effect of the
soft gluon emission is that it is possible to compute radiative
corrections in momentum space expanding $S$, $\tilde{S}$,
$\overline{S}$, $\tilde{\overline{S}}$ and their
hermitian conjugates in powers of $g$, while previous approaches
performed the radiative corrections mainly in coordinate space by
expanding $S(x)$, $\tilde{S} (x)$, $\overline{S} (x)$,
$\tilde{\overline{S}} (x)$ which are the Fourier transforms of $S$,
$\tilde{S}$, $\overline{S} $, $\tilde{\overline{S}}$. In
Ref.~\cite{korchem87}, the authors computed the cusp anomalous
dimension in the momentum space. The study of the analytic structure
on the soft Wilson lines can be extended to the collinear Wilson line
and the processes in which the different analytic structure of the
collinear Wilson lines may affect the scaling behavior are under
investigation. 

Finally, let us comment on the approaches using the forward scattering
amplitudes and the matrix element squared, which should give the same
physical results on the radiative corrections due to the optical
theorem. The goal of SCET is to obtain the effective operators in
$\mathrm{SCET}_{\mathrm{II}}$ from the time-ordered products of the
currents and take the matrix elements of the operators. The approach
using the forward scattering amplitude fits this purpose.
In the approach using the matrix elements squared, it is possible to
compute radiative corrections, but the final result cannot be
expressed in terms of operators. And the computation is more involved
since we should consider the Feynman rules for the propagators with
and without the cuts, and we should also take into account which
particles are on which side of the cut. Though it is formidable, it
has been done. In this approach also, we can draw the
soft Wilson lines in spacetime. Due to the cancellation of the
soft Wilson lines, the overall soft Wilson line becomes a connected
line, and it is easy to see which cusp angles contribute. This is
seen, for example, in Refs.~\cite{Korchemsky:1992xv,
  Bauer:2002ie}. However, care should be taken in obtaining the final
result. In deep inelastic scattering in Ref.~\cite{Korchemsky:1992xv},
there are contributions from two cusp angles, but the contribution
of the cusp angle from the complex-conjugated amplitude should be
changed to the cusp angle from the original amplitude by taking the 
hermitian conjugate because it is  on the opposite side of
the cut. This corresponds to taking the same cusp angle in the
original part, not the cusp angle on the opposite side of the
cut. This also happens in $e^+e^-$ collisions if we take the path of
the soft Wilson line given in Ref.~\cite{Bauer:2002ie}, and we have to
consider choosing the correct cusp angle as in the case of the deep
inelastic scattering. On the other hand, we can draw the soft Wilson
lines in a similar way in the approach using the forward scattering
amplitudes, but with straightforward rules. As we have explained, we
can specify the configurations of the soft Wilson lines for a given 
physical process and compute the effects of the soft gluon emission in
a straightforward manner and can see which cusp angles contribute.

\section*{Acknowledgments}
J.~Chay was supported by Korea Research Foundation Grant
(KRF-2003-041-C20052), and by Korea University. C.~Kim was supported
by the National Science Foundation under Grant
No. PHY-0244599. J.~P.~Lee was supported by the BK21 Program of the
Korean Ministry of Education. Y.~G.~Kim was supported by the Korean
Federation of Science and Technology Societies through the Brain Pool
program.

\end{document}